\documentclass[useAMS,usenatbib]{mn2e}
\usepackage{color}
\usepackage{graphicx}
\usepackage{rotating}
\usepackage{lscape}

\title[Physical parameters of binary stars]
{Physical parameters of 62 eclipsing binary stars using the ASAS-3 data$-$I}
\author[Deb \& Singh]{Sukanta Deb\thanks{E-mail:
sdeb@physics.du.ac.in; sukantodeb@gmail.com}, Harinder P. Singh\\
Department of Physics \& Astrophysics, University of Delhi,
         Delhi 110007, India\\
}
\begin{document}

\date{Received on ; Accepted on }

\pagerange{\pageref{firstpage}--\pageref{lastpage}} \pubyear{2010}

\maketitle

\label{firstpage}

\begin{abstract}
We present a detailed light curve analysis of publicly 
available V band observations of 62 binary stars, mostly contact binaries, 
obtained by the All Sky Automated Survey (ASAS)-3 project between 
2000 and 2009. Eclipsing binaries are  important astronomical targets for 
determining the physical parameters of component stars from the geometry. 
They provide an independent direct method of measuring the radii of stars. 
We improved the ASAS determined periods, ephemeris and obtained 
the Fourier parameters from the phased light curves of these 62 stars. These 
Fourier parameters were used for preliminary classification of the stars in 
our sample. The phased light curves were then analysed with the aid of the 
Wilson-Devinney light curve modelling technique in order to obtain various 
geometrical and physical parameters of these binaries. The spectroscopic mass 
ratios as determined from the the radial velocity measurements available in 
the literature were used as one of the inputs to the light curve modelling. 
Thus reliable estimations of parameters of these binaries were obtained with 
combined  photometric and spectroscopic data and error estimates were made 
using the heuristic scan method. For several systems in the 
sample, the solutions were obtained for the first time and would serve as a
good source in the future for the light curve analysis based on the more 
precise follow-up CCD photometric observations. Out of 62 stars in the sample, 
photometric analysis of 39 stars are presented here for the first time using 
the ASAS photometry and precise spectroscopic mass ratios. From the analysis, 
we found 54 contact binaries, 6 semi-detached binaries and 2 detached binaries.
The Fourier parameters in the a$_{2}-$a$_{4}$ plane were used for preliminary 
classification and the final classification was done based on the Roche lobe 
geometry obtained from the light curve modelling.    
\end{abstract}

\section{Introduction}

Eclipsing binaries play an important astrophysical tool for the study of 
star formation, stellar structure, in testing the theories of stellar evolution
 and determining the physical properties of stars. They have been used as 
standard candles \citep{pacz97} to determine the size and structure of the 
Galaxy, as well as to constrain the cosmological distance ladder \citep{bona06}.
They are the potential targets since the orbital motion, inferred from the 
radial velocity curves and the shape of eclipses from the photometric light 
curves can be entirely explained by the gravitation laws and geometry of the 
system. They are also useful for distance determinations since the radii and
temperatures of the components can be reliably determined from the combined 
photometry and spectroscopy \citep{vila10}. 

Contact binaries, also known as the W UMa stars, are low mass eclipsing 
binaries consisting of ellipsoidal components with orbital periods less than
1 day, usually in the range $0.2< P <0.8 $ day, with continuously changing 
brightness \citep{kang02, perc07}. They consist of solar type 
main-sequence components which fill their Roche lobes and share a common outer 
envelope at the inner Lagrangian point. The maxima of the light 
curves are rounded as the shapes of the two stars are strongly tidally 
distorted because of their proximity to one another \citep{shu82}. Most of the 
W UMa type binaries have temperatures of the components roughly equal because 
of sharing a common envelope with the same entropy, thereby making the 
effective temperatures almost equal over the surfaces \citep{pacz06}. 
There is also a subclass of contact binaries with temperature difference of 
1000~K or larger between the components. \citet{csiz04} called these as B-type 
contact binaries. In this paper, we follow the definition of B-type contact 
binaries as given by \citet{csiz04}. B-type contact binaries are sometimes 
called `Poor thermal contact binaries' \citep{ruci_du}. 
     
The ASAS project started in 1997 with the goal of photometrically monitoring 
millions of stars brighter than 14 mag in the V band and distributed all over 
the sky \citep{pojm97, pojm02}. The ASAS-3 system is located at Las Campanas
Observatory and is operating since August 2000. It is equipped with 200/2.8 
Minolta telephoto lenses and covers 8.8 $\times$ 8.8 deg of the sky through the
 standard V and I filters. So far, ASAS has discovered more than 50000 
variables located south of declination $+28^{o}$ and nearly 10000, among these 
newly identified variables, are eclipsing binaries. For more details about the 
ASAS project and the instruments, we refer the reader to \citet{pojm97,pojm98,
pojm02}. These newly discovered binaries, among the others, are very valuable 
and their determination of absolute physical parameters will help to address 
several key problems related to stellar astrophysics.

ASAS database has time series photometric measurements of a large number of 
eclipsing binary stars.  The light curves of these stars can provide reliable 
information about various parameters, once the mass ratios from the radial 
velocity measurements are available. The evolution of stars with cool 
convective envelopes; notably, of binary stars, are greatly affected by the 
dynamo-driven mass loss, magnetic braking and tidal friction which cause 
angular momentum loss from the orbit \citep{pari09}. Therefore, enriching the 
total sample of new eclipsing binary stars with their parameters preliminary 
determined from databases like ASAS and their future follow-up regular 
observations using more precise CCD photometric data and spectroscopic radial 
velocity measurements will enhance our knowledge in understanding the nature, 
structure and evolution of binary stars. 

Recently, ASAS photometric database has been used by \citet{helm09} to 
determine the absolute and physical parameters for a sample of 18 detached 
binaries using the spectroscopic values of mass ratio $q_{\rm sp}$ obtained 
from their new radial velocity measurements. Out of 18 
eclipsing binaries in their sample, 15 are new, discovered by ASAS. 

The main rationale of this paper is to exploit the available ASAS photometric 
data of 62 binary stars which have complementary well-defined radial velocity 
measurements in the literature in order to determine their geometrical and 
physical parameters. In Section~\ref{sample}, we describe the criteria to 
select the sample in our present study. In Section~\ref{sect_peri}, we 
describe a method to improve the period as determined by ASAS. 
Section~\ref{sect_clas} deals with the determination of Fourier parameters 
from the phased light curves and classification using the Fourier coefficients 
in the a$_{2}-$a$_{4}$ plane. In Section~\ref{sect_mode}, we describe the light 
curve analysis using the WD light curve modelling technique to estimate the 
various geometrical and physical parameters of the 62 ASAS stars. In 
Section~\ref{sect_resu}, we present results obtained from the light curve 
modelling and compare to those in the literature, wherever available. 
Section~\ref{sect_rela} describes various relations for contact binaries. 
Finally, in Section~\ref{summary},  we present summary and conclusions of this 
study. 
\section{Sample selection criteria}
\label{sample}
\subsection{The data}
The number of photometric light curves of eclipsing binary stars have 
surpassed their spectroscopic radial velocity measurements with the 
advent of various all sky automated photometric surveys. Rucinski \& his
collaborators \cite[hereafter RC,][]{ruci1,ruci2,ruci3,ruci4,ruci5,ruci6,ruci7,ruci8,ruci9,ruci10,ruci11,ruci2sp,ruci1wi,ruci12,ruci13,ruci14,ruci15} started a 
project to obtain precise radial velocity measurements of a large number of 
binary stars. In practice, physical parameters of eclipsing binary stars are 
derived from the photometric light curve modelling which in general, require 
accurate and precise  mass ratio as an input obtained from the spectroscopic 
radial velocity measurements. 

We searched for eclipsing binaries from the ASAS online database, which have 
well-defined determinations of mass ratios available from radial velocity 
measurements obtained by RC. We found 67 stars in the ASAS database 
corresponding to the coordinates of identifiers of stars in the RC's sample 
using the SIMBAD database. However, light curves of the 5 ASAS stars turned 
out to be very noisy after they were phased  with  their respective periods as 
given in the database. The light curves of these stars could not even be 
improved using the period determination technique as 
described in Section~\ref{sect_peri}. Therefore, these stars were rejected for 
further analysis. These stars were 002524$-$4655.5 (BQ Phe), 050702$-$0047.6 
(V1363 Ori), 124015$-$1848.0 (SX Crv),  131223$+$0239.3 (KZ Vir) and 
164824$+$1708.1 (V918 Her). Our final sample, thus, consisted of 62 eclipsing 
binary stars.  Light curves of each of the 62 stars were then examined 
carefully and outliers were removed after visual inspection. All the 
selected targets have evenly covered V-band light curves with an internal 
accuracy $\sim 0.03-0.11$ mag having number of data points between 96 
(062605$+$2759.9) and 1221 (204628$-$7157.0). The data for these 62 eclipsing 
binary stars, mostly contact binaries from the ASAS-3 database along with 
their spectroscopically determined mass ratios are used to model the light 
curves in order to obtain their various geometrical and physical parameters. 
The parameters determined for these 62 stars in this work, in turn, can be 
used as inputs to the analysis of more accurate and precise light curves based 
on follow-up observations. All the systems have their ASAS photometry 
available online \footnote{http://www.astrouw.edu.pl/asas}. On the 
average, the number of observations per star is of the order of several 
hundreds in the ASAS database. These stars have a good phase coverage of their
light curves.  
\subsection{Radial velocity and its importance}       
It is a widely accepted fact that physical parameters of binary stars obtained 
from the photometric light curve modelling are reliable only when the 
spectroscopically determined mass ratio is used as an input in the photometric 
light curve modelling and kept fixed. In the light curve modelling of binary 
stars, mass ratio is the most important parameter and should be obtained from 
precise spectroscopic radial velocity measurements. It plays an important 
role in deriving the accurate masses and radii of the components, which are 
the key parameters in understanding the structure and evolution of binary 
stars. It has been seen from the literature that determination of mass ratio 
from the photometric light curve modelling is exceedingly difficult. For a 
partially-eclipsing binary, it is impossible to reliably measure the mass 
ratio from the photometric modelling even with the highly accurate photometric 
data. On the other hand, the mass ratios determined from the photometric data 
may be reliable only for a total-eclipse configuration of the binary systems 
as such light curves show characteristic inner contacts with duration of 
totality setting a strong constraint on the $(q,i)$ pair \citep{ruci_q}. The 
literature is crowded with physical parameter estimations using the photometric
 mass ratio ($q_{\rm ph}$) obtained from the light curve modelling for many 
partially eclipsing systems which were later modified when subsequent radial 
velocities became available.      
\section{Updating the ASAS period}
\label{sect_peri}
The automated surveys like ASAS, offer a unique opportunity to study the 
properties of a large number of stars. However, the processing of huge amount 
of data in these databases is quite challenging, even when looking at 
seemingly small issues such as period determination and classification 
\citep{dere07}. Although the periods determined for most of the stars in the 
ASAS database is reasonably accurate (except 185318$+$2113.5), 
we have found that the ASAS periods need to be updated to the 
improved quality of data available at present. Periods of variable stars 
in the ASAS database were in general determined very well during the 
preparation of the catalog. However, the data for some parts of the sky were 
analysed many years ago, when the time span was much shorter. Therefore, the 
accuracy was not good as it can be obtained today.

It has been found that the period determination can be further improved by the 
use of methods such as string lengths \citep{dere07} or minimization of entropy
 method \citep{deb_basi}. The new period determinations result in the 
improvement of the  ASAS periods to fifth or sixth decimal place. This period 
improvement has resulted in revealing the accurate shapes of the phased light 
curves of the ASAS eclipsing binaries. In practice, when the period is accurate
 enough, the phased data are clearly arranged with minimum scatter due to noise.
 On the other hand, when the light curve is folded into phase by an incorrect 
period, the data gets randomized and more scattered. 

\begin{figure}
\begin{center}
\includegraphics[height=9cm,width=9cm]{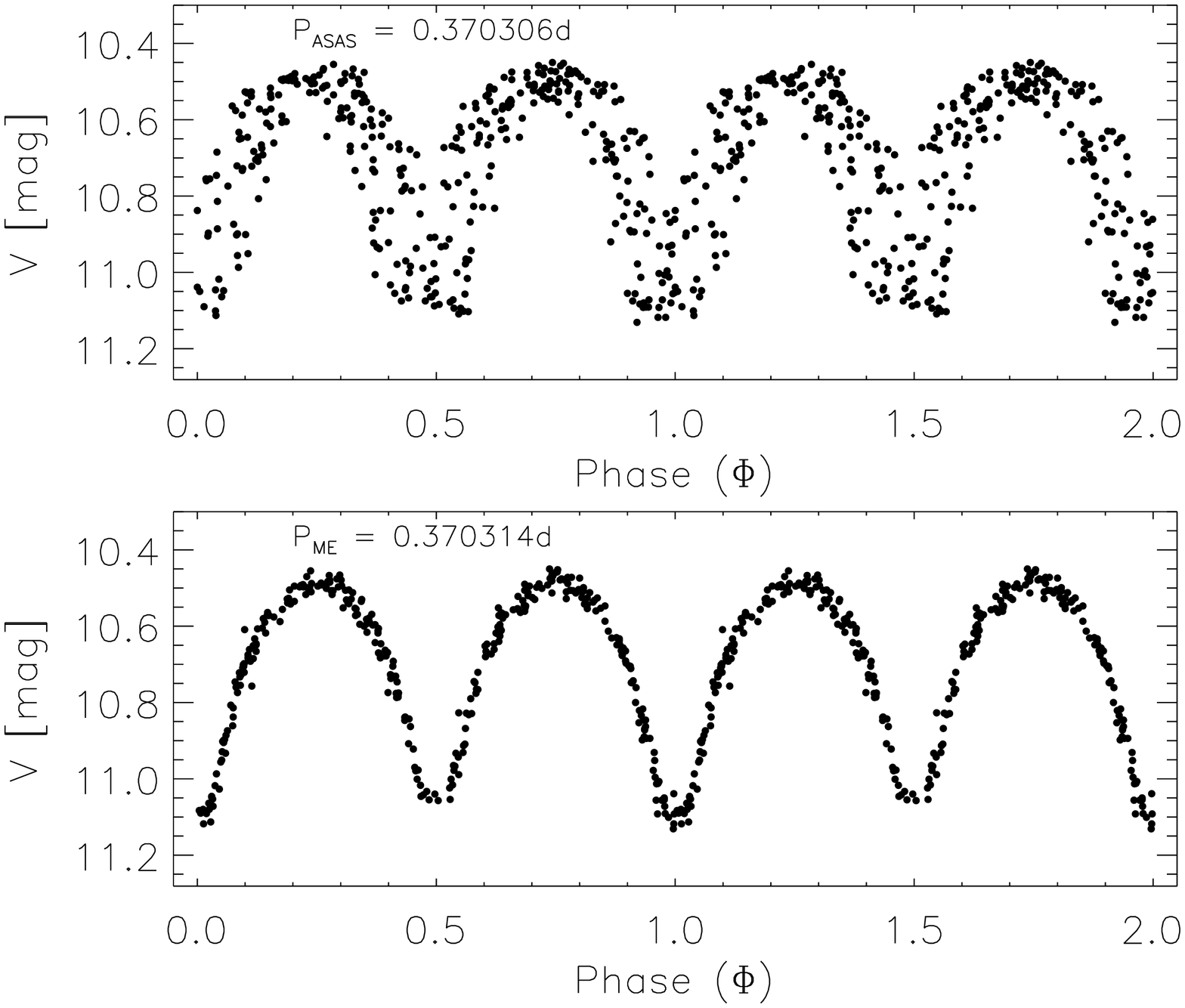}
\end{center}
\caption{An example (193524$+$0550.3) of improvements in the ASAS period (top)
determined by using the minimization of entropy (bottom) method.}
\label{lc_imp}
\end{figure}

\begin{figure}
\begin{center}
\includegraphics[height=10cm,width=9cm]{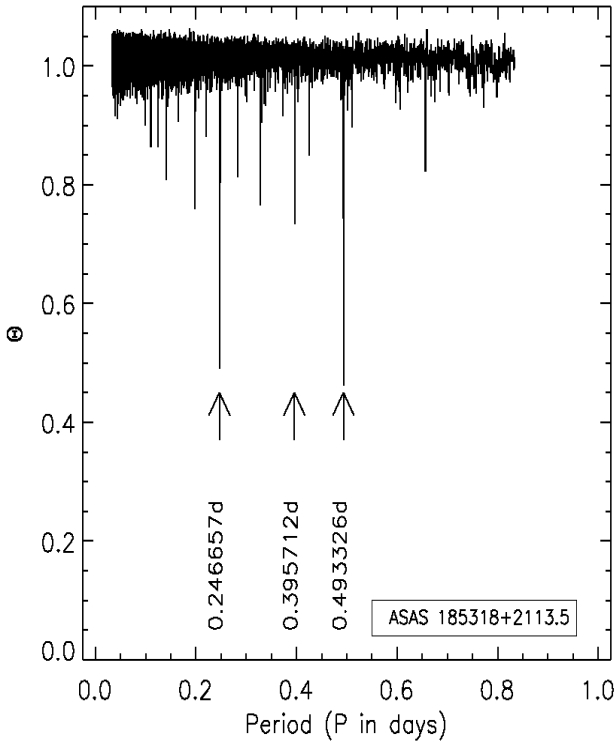}
\end{center}
\caption{Plot of $\Theta$ versus period for ASAS 185318+2113.5 as determined 
from PDM.}
\label{v1003_her1}
\end{figure}

\begin{figure}
\begin{center}
\includegraphics[height=10cm,width=9cm]{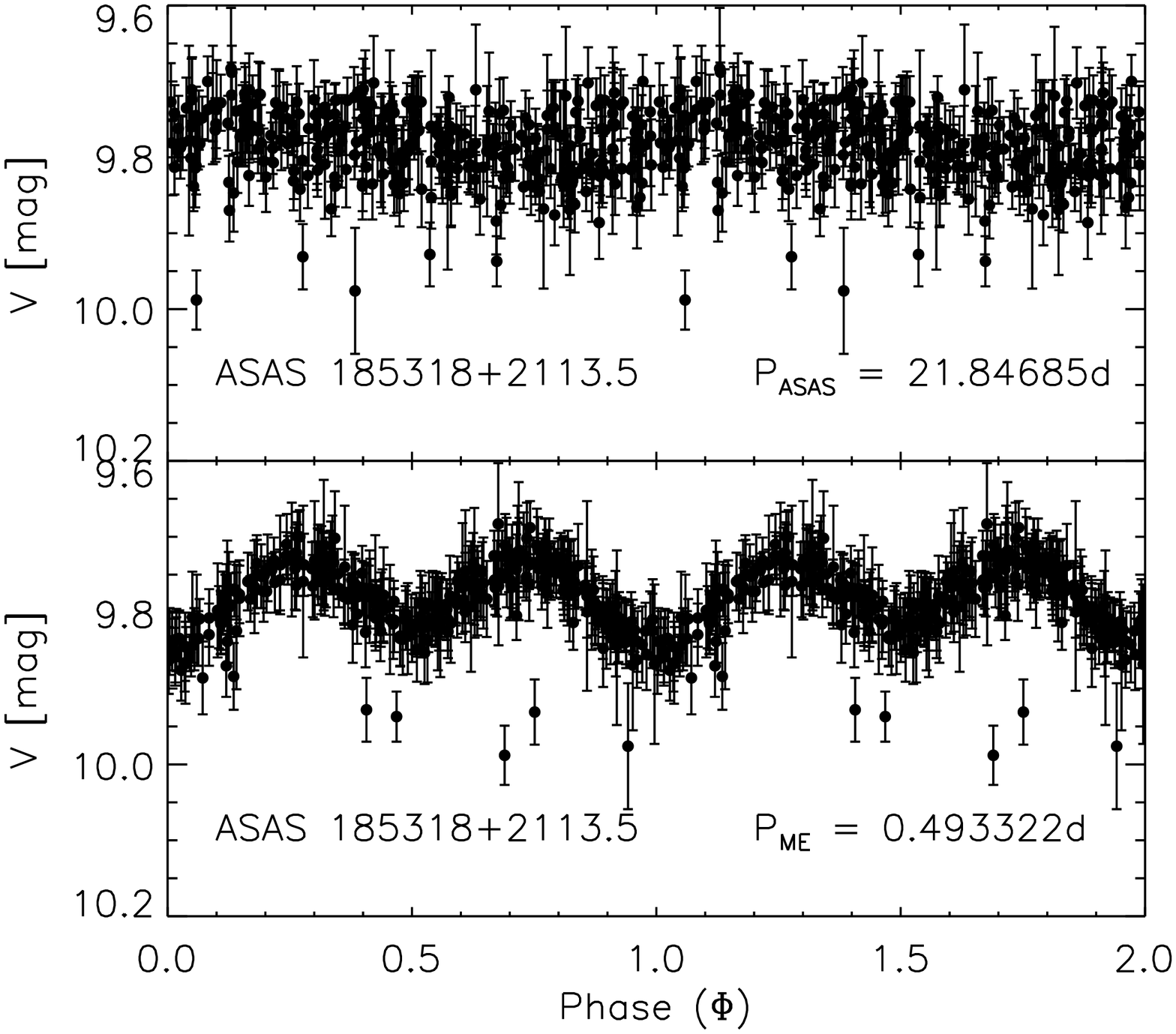}
\end{center}
\caption{The upper panel shows the phased light curve of ASAS 185318+2113.5 
determined from the ASAS period P$_{\rm ASAS}$ = 21.84685d. The lower panel 
shows the phased light curve of the same star with period P$_{\rm ME}$ = 0.493322d.}
\label{v1003_her2}
\end{figure}

We have used the minimization of entropy (hereafter, ME) method to improve and 
update the period as determined by ASAS which is based on the minimization of the information entropy of the light curve \citep{cinc95}. It is based on the 
fact that the light curves in the phase-magnitude ($\Phi$, x) space are more ordered when the phases are determined using the correct period, where phase ($\Phi$) is defined as:
\begin{eqnarray}
\Phi = \frac{\left( t-t_{0}\right) }{P}-Int\left( \frac{\left( t-t_{0}\right) }{P}\right), 
\end{eqnarray}
where $t_{0}$ is the initial epoch of the variable star light curve and $t$ is 
the time of observations. The value of $\Phi$ is from 0 to 1, corresponding 
to a full cycle of the orbital period $P$. $Int$ denotes the integer part of 
the quantity. The light curve data ($t_{i},x_{i}$) is phased into the 
phase-magnitude space according to the trial period. Now the phase-magnitude 
space is divided into  $n\times m$ bins. Let $\mu_{ij}$ be the number of 
observations in the $(i,j)$ bin normalized by the total number of the data, 
then the entropy $S$ is 
\begin{eqnarray}
S=\sum_{i=1}^{n}\sum_{j=1}^{m}\mu_{ij}ln(\mu_{ij}) , 
\end{eqnarray}    
for all $\mu_{ij}$ $>$ 0.
If the trial period is not the true one, the light curve points will be 
distributed uniformly in the ($\Phi, x$) plane and the entropy will be 
maximum. On the other hand, if the trial period matches with the actual 
period, the points will be limited into a small regions of the plane and 
the entropy will then be minimum, which will correspond to the best period 
of the light curve. In our case, we applied the ME method around $\pm1$\% of 
the ASAS-determined periods. An example of the period improvement is shown in 
Fig.~\ref{lc_imp}. The typical improvement results in a change in the fifth or 
sixth decimal place. For 185318$+$2113.5, we have found that the ASAS period 
is not correct, as no significant periodicity can be confirmed from the 
phased light curve with P$_{\rm ASAS} = 21.84685$ days. For preliminary period
determination of this star only, we have used the phase dispersion minimization method 
\citep[hereafter PDM,][]{stel78}. In Fig.~\ref{v1003_her1}, we show the PDM 
results applied to the light curve of ASAS 185318$+$2113.5. The three dominant 
peaks of the PDM results are 0.493326d, 0.246658d and 0.395712d in increasing 
$\Theta$ values, where $\Theta$ is defined as in \citet{stel78}. We also tried to improve the most dominant period 0.493326d, using the ME 
method. ME method is run around $\pm$1\% of the period 0.493326d. The period 
is found to be P$_{\rm ME} = 0.493322$d. In this case, the period determined 
by both PDM and ME methods are nearly identical. The light curve of ASAS 
185318$+$2113.5 phased with the ASAS determined period P$_{\rm ASAS}$ and 
our determined period P$_{\rm ME}$ are shown in Fig.~\ref{v1003_her2}. Our 
photometric analysis reveals that the star is at a very low inclination angle 
$i[^{o}] = 41.59 \pm 2.78 $ and is a B-type contact binary with 
$\Delta T = 2316 \pm 647$~K. However, little noisiness of the ASAS-3 data and 
its low amplitude makes this star a good candidate for accurate follow-up 
observations.   
    
 In Table~\ref{basic_info}, we list the serial numbers, ASAS identification 
names, SIMBAD identification names, J2000 co-ordinates, 
ASAS Periods (P$_{\rm ASAS}$), updated periods determined using 
the ME method (P$_{\rm ME}$), type of binary determined from the light curve 
analysis, V band magnitudes, infrared $(J-K)$ color indices \citep{cutr03} and 
epochs of minimum light determined from the light curve modelling. In 
Table~\ref{radial_info}, we list the spectral type, radial velocity 
semi-amplitudes of the components and their errors, mass ratios and references 
for the radial velocity measurements.    
\begin{table*}
\begin{center}
\caption{Basic information of the ASAS stars used used for the analysis}
\label{basic_info}
\scalebox{0.9}{
\begin{tabular}{cclcclcllcc}
\hline
Serial No.&ASAS ID & Other ID & $\alpha (2000)$ & $\delta (2000)$ &P$_{\rm ASAS}$~[d]& P$_{\rm ME}$~[d] &Type & V~[mag] & J-K~[mag] & Epoch(PHOEBE) \\ 
&&&&&&&&&&(JD$-$2450000) \\ \hline
 1&003628$+$2132.3 &  DZ Psc	&      00 36 27.95 &	 $+$21 32 14.63 & 0.366125 &0.366131  &   A    &  11.126  & 0.253(35)  & 2970.55742(53) \\
 2&011638$-$3942.S &  AD Phe	&      01 16 38.07 &	 $-$39 42 31.32 & 0.37992  &0.379919  &   W    &  10.290  & 0.400(33)  & 1903.57897(27) \\
 3&012104$+$0736.3 &  AQ Psc	&      01 21 03.56 &	 $+$07 36 21.63 & 0.475611 &0.475604  &   A    &  8.67    & 0.255(30)  & 3653.71608(47) \\
 4&014656$-$0945.1 &  TT Cet	&      01 46 56.50 &	 $-$09 45 09.74 & 0.48596  &0.485952  &   B    &  10.9    & 0.250(29)  & 5103.85398(27) \\
 5&014854$-$2053.6 &  TW Cet	&      01 48 54.14 &	 $-$20 53 34.60 & 0.316849 &0.316851  &   A  &    10.7    & 0.492(38)  & 4476.61655(09) \\
 6&023833$-$1417.9 &  DY Cet	&      02 38 33.18 &	 $-$14 17 56.73 & 0.440790 &0.440790  &   A    &  9.8	  & 0.258(34)  & 3644.73772(18) \\
 7&024952$+$0856.3 &  EE Cet	&      02 49 52.26 &	 $+$08 56 17.95 & 0.37992  &0.379925  &   W    &  9.6	  & 0.280(47)  & 2643.60106(42) \\
 8&030701$-$5608.1 &  SZ Hor	&      03 07 01.23 &	 $-$56 08 04.45 & 0.62511  &0.625102  &   B    &  11.214  & 0.217(31)  & 4314.86179(21) \\
 9&030953$-$0653.6 &  UX Eri	&      03 09 52.74 &	 $-$06 53 33.56 & 0.44528  &0.445289  &   W    &  10.639  & 0.374(33)  & 2954.68457(31) \\
10&033459$+$1742.6 &  V1123 Tau &      03 34 58.54 &	 $+$17 42 38.04 & 0.399946 &0.399947  &   W    &  9.97    & 0.246(50)  & 2621.64645(28) \\
11&034814$+$2218.9 &  EQ Tau	&      03 48 13.43 &	 $+$22 18 50.92 & 0.341347 &0.341350  &   A    &  11.196  & 0.376(28)  & 4846.56462(27) \\
12&034928$+$1254.7 &  V1128 Tau &      03 49 27.77 &	 $+$12 54 43.84 & 0.305372 &0.305371  &   W	& 9.353   & 0.454(30)  & 3751.55758(13) \\
13&035153$-$1031.8 &  BV Eri	&      03 51 53.34 &	 $-$10 31 49.46 & 0.507654 &0.507655  &   B    &  8.32    & 0.221(33)  & 4300.91691(25) \\
14&041209$-$1028.2 &  YY Eri	&      04 12 08.85 &	 $-$10 28 09.96 & 0.321499 &0.321498  &   W    &  8.41    & 0.443(34)  & 3454.51668(15) \\
15&042925$-$3334.6 &  CT Eri	&      04 29 24.79 &	 $-$33 34 34.33 & 0.63419  &0.634196  &   B    &  10.22   & 0.215(36)  & 5113.78813(36) \\
16&051114$-$0833.4 &  ER Ori	&      05 11 14.50 &	 $-$08 33 24.66 & 0.4234   &0.423406  &   W    &  10.00   & 0.317(31)  & 1967.52536(27) \\
17&051832$-$6813.6 &  RW Dor	&      05 18 32.54 &	 $-$68 13 32.76 & 0.285461 &0.285463  &   W    &  10.80   & 0.573(33)  & 3466.52950(11) \\
18&062605$+$2759.9 &  AH Aur	&      06 26 04.93 &	 $+$27 59 56.40 & 0.247053 &0.494106  &   A    &  10.176  & 0.329(28)  & 2970.78867(66) \\
19&064558$-$0017.5 &  DD Mon	&      06 45 57.83 &	 $-$00 17 31.90 & 0.56803  &0.568019  &   ESD  &  11.1    & 0.388(35)  & 3051.72047(22) \\
20&071058$-$0352.8 &  V753 Mon  &      07 10 57.85 &	 $-$03 52 43.19 & 0.67705  &0.677045  &   ESD  &  8.34    & 0.203(32)  & 2701.61206(17) \\
21&073246$-$2047.5 &  TY Pup	&      07 32 46.23 &	 $-$20 47 31.07 & 0.81925  &0.819243  &   A    &  8.62    & 0.257(61)  & 1876.71731(25) \\
22&073338$-$5007.4 &  HI Pup	&      07 33 38.21 &	 $-$50 07 25.05 & 0.432618 &0.432618  &   A    &  10.70   & 0.349(30)  & 1924.66065(40) \\
23&073905$-$0239.1 &  V868 Mon  &      07 39 04.83 &	 $-$02 39 05.54 & 0.637704 &0.637704  &   A    &  8.9	  & 0.160(33)  & 3745.73825(63) \\
24&084002$+$1900.0 &  TX Cnc	&      08 40 01.71 &	 $+$18 59 59.51 & 0.382882 &0.382883  &   W    &  10.026  & 0.355(27)  & 4409.86331(45) \\
25&084108$-$3212.1 &  TZ Pyx	&      08 41 08.26 &	 $-$32 12 03.02 & 2.318500 &2.318530  &   ED   &  10.67   & 0.083(33)  & 3805.78546(51) \\
26&100141$+$1724.5 &  XY Leo	&      10 01 40.43 &	 $+$17 24 32.70 & 0.284098 &0.284101  &   W    &  9.67    & 0.701(41)  & 2733.60478(28) \\
27&100234$+$1702.8 &  XZ Leo	&      10 02 34.19 &	 $+$17 02 47.14 & 0.487736 &0.487736  &   A    &  10.29   & 0.171(30)  & 3339.81443(30) \\
28&100248$+$0105.7 &  Y Sex	&      10 02 47.96 &	 $+$01 05 40.35 & 0.41982  &0.487736  &   A    &  9.97    & 0.278(32)  & 4589.56751(21) \\
29&101602$-$0618.5 &  XX Sex	&      10 02 47.96 &	 $+$01 05 40.35 & 0.540111 &0.540108  &   A    &  9.32    & 0.243(38)  & 4164.67601(40) \\
30&104033$+$1334.0 &  UZ Leo	&      10 40 33.19 &	 $+$13 34 00.86 & 0.61806  &0.618057  &   A    &  9.75    & 0.208(30)  & 3809.70840(26) \\
31&105030$-$0241.7 &  VY Sex	&      10 50 29.72 &	 $-$02 41 43.05 & 0.443432 &0.443433  &   W    &  9.01    & 0.346(38)  & 4918.58983(21) \\
32&110211$+$0953.7 &  AM Leo	&      11 02 10.89 &	 $+$09 53 42.69 & 0.365798 &0.365799  &   W    &  9.31    & 0.328(30)  & 3739.83444(20) \\
33&110505$+$0509.1 &  AP Leo	&      11 05 05.02 &	 $+$05 09 06.42 & 0.430356 &0.430358  &   A    &  9.53    & 0.368(33)  & 4656.47307(33) \\
34&120103$+$1300.5 &  AG Vir	&      12 01 03.50 &	 $+$13 00 30.02 & 0.64265  &0.642648  &   A    &  8.52    & 0.145(32)  & 3190.48336(47) \\
35&121206$+$2232.0 &  CC Com	&      12 12 06.20 &	 $+$22 31 58.00 & 0.220686 &0.220686  &   W    &  11.00   & 0.741(26)  & 3012.86303(16) \\
36&123300$+$2642.9 &  RW Com	&      12 33 00.28 &	 $+$26 42 58.38 & 0.237348 &0.237346  &   W    &  11.00   & 0.618(34)  & 4918.69861(20) \\
37&131032$-$0409.5 &  PY Vir	&      13 10 32.22 &	 $-$04 09 32.59 & 0.311251 &0.311248  &   ESD  &  9.60    & 0.572(38)  & 5002.62738(25) \\
38&134607$+$0506.9 &  HT Vir	&      13 46 06.75 &	 $+$05 06 56.27 & 0.407672 &0.407672  &   ESD  &  7.16    & 0.324(30)  & 4919.78397(20) \\
39&141726$+$1234.1 &  VW boo	&      14 17 26.03 &	 $+$12 34 03.45 & 0.342315 &0.342315  &   A    &  10.539  & 0.467(29)  & 2840.61141(20) \\
40&141937$+$0553.8 &  NN Vir	&      14 19 37.74 &	 $+$05 53 46.66 & 0.480687 &0.480688  &   A    &  7.64    & 0.240(37)  & 3576.56765(36) \\
41&143504$+$0906.8 &  CK Boo	&      14 35 03.76 &	 $+$09 06 49.39 & 0.355154 &0.355152  &   A    &  9.09    & 0.295(44)  & 2670.83810(29) \\
42&144803$+$1356.7 &  EL Boo	&      14 48 03.40 &	 $+$13 56 41.19 & 0.413767 &0.413766  &   ESD  &  9.35    & 0.282(34)  & 4633.62388(64) \\
43&152243$+$1615.7 &  OU Ser	&      15 22 43.47 &	 $+$16 15 40.74 & 0.148382 &0.296768  &   A    &  8.25    & 0.376(31)  & 4891.85575(51) \\
44&153152$-$1541.1 &  VZ Lib	&      15 31 51.76 &	 $-$15 41 10.19 & 0.358259 &0.358256  &   A    &  10.341  & 0.379(35)  & 4487.86967(22) \\
45&155649$+$2216.0 &  AU Ser	&      15 56 49.40 &	 $+$22 16 01.00 & 0.386498 &0.386498  &   A    &  10.50   & 0.509(28)  & 4299.48288(29) \\
46&164121$+$0030.4 &  V502 Oph  &      16 41 20.86 &	 $+$00 30 27.37 & 0.45339  &0.453388  &   W    &  8.53    & 0.345(29)  & 4701.54142(29) \\
47&165717$+$1059.8 &  V2357 Oph &      16 57 16.76 &	 $+$10 59 51.38 & 0.207783 &0.415568  &   A    &  10.52   & 0.426(31)  & 4691.59160(93) \\
48&171358$+$1621.0 &  AK Her	&      17 13 57.82 &	 $+$16 21 00.61 & 0.421522 &0.421524  &   A    &  8.51    & 0.313(34)  & 5071.56586(24) \\
49&173356$+$0810.0 &  V2377 Oph &      17 33 56.05 &	 $+$08 09 57.81 & 0.425403 &0.425406  &   W    &  8.56    & 0.333(34)  & 3652.52046(66) \\
50&175332$-$0354.9 &  V2610 Oph &      17 53 32.26 &	 $-$03 54 55.33 & 0.42652  &0.426514  &   ESD  &  9.20    & 0.393(34)  & 2508.56035(42) \\
51&180921$+$0909.1 &  V839 Oph  &      18 09 21.27 &	 $+$09 09 03.63 & 0.409    &0.409008  &   A    &  8.984   & 0.344(28)  & 3547.70359(15) \\
52&182913$+$0647.3 &  V2612 Oph &      18 29 13.01 &	 $+$06 47 13.72 & 0.3753   &0.375309  &   W    &  9.44    & 0.279(38)  & 4234.79641(40) \\
53&185318$+$2113.5 &  V1003 Her &      18 53 17.54 &	 $+$21 13 32.74 & 21.84685 &0.493322   &   B    & 9.795   & 0.234(32)  & 3099.90625(96) \\
54&193524$+$0550.3 &  V417 Aql  &      19 35 24.12 &	 $+$05 50 17.66 & 0.370306 &0.370314  &   W    &  10.647  & 0.354(33)  & 4216.88607(13) \\
55&194813$+$0918.5 &  OO Aql	&      19 48 12.65 &	 $+$09 18 32.38 & 0.506786 &0.506794  &   A    &  9.51    & 0.440(36)  & 5070.71438(18) \\
56&203113$+$0513.2 &  MR Del	&      20 31 13.47 &	 $+$05 13 08.50 & 0.52169  &0.521692  &   ED   &  11.01   & 0.624(28)  & 4742.53633(35) \\
57&204628$-$7157.0 &  MW Pav	&      20 46 27.76 &	 $-$71 56 58.48 & 0.795    &0.794994  &   A    &  8.80    & 0.229(30)  & 1887.51161(26) \\
58&205710$+$1939.0 &  LS Del	&      20 57 10.29 &	 $+$19 38 59.22 & 0.36384  &0.363842  &   W    &  8.654   & 0.308(37)  & 4628.80678(97) \\
59&222257$+$1619.4 &  BB Peg	&      22 22 56.89 &	 $+$16 19 27.84 & 0.361487 &0.361502  &   W    &  10.967  & 0.388(28)  & 3586.70831(18) \\
60&233655$+$1548.1 &  V407 Peg  &      23 36 55.37 &	 $+$15 48 06.43 & 0.63688  &0.636882  &   A    &  9.28    & 0.181(33)  & 5084.67263(51) \\
61&234535$+$2528.3 &  V357 Peg  &      23 45 35.06 &	 $+$25 28 18.94 & 0.57845  &0.578450  &   A    &  9.06    & 0.178(34)  & 4693.78602(34) \\
62&234718$-$0805.2 &  EL Aqr	&      23 47 18.35 &	 $-$08 05 12.09 & 0.48141  &0.481412  &   A    &  10.572  & 0.249(35)  & 4456.57167(35) \\
 		 
\hline
\end{tabular}}
\end{center}
\end{table*}
\begin{table*}
\begin{center}
\caption{Spectral type and radial velocity information for the ASAS stars}
\label{radial_info}
\scalebox{0.9}{
\begin{tabular}{cllllllc}
\hline
ASAS ID & Spectral type & $K_{1}$ [km~s$^{-1}$] &  $\sigma K_{1}$ [km~s$^{-1}$]& $K_{2}$ [km~s$^{-1}$] &  $\sigma K_{2}$ [km~s$^{-1}$] &Mass ratio ($q$)&References \\\hline
003628$+$2132.3&       F7V&     40.39&      2.64&    297.98&      4.18&0.136(10)& 7 \\
011638$-$3942.S&    F9/G0V&     89.04&      3.10&    242.41&      1.42&0.370(10)&11 \\
012104$+$0736.3&     F5/8V&     59.50&      0.40&    263.30&      1.90&0.226(02)& 1, 13 \\
014656$-$0945.1&       F4V&    107.90&     10.30&    276.00&     27.00& 0.39(07)&11 \\
014854$-$2053.6&       G1V&    157.40&      3.26&    209.66&      3.84& 0.75(03)&11 \\
023833$-$1417.9&       F5V&     90.84&      2.11&    255.40&      2.28& 0.356(9)&13 \\
024952$+$0856.3&       F8V&     84.05&      1.24&    266.92&      1.54& 0.315(5)& 6 \\
030701$-$5608.1&       F3V&    109.51&      6.60&    231.34&     13.50& 0.47(04)&11 \\
030953$-$0653.6&       F9V&     91.75&      1.55&    245.76&      1.86&0.373(21)&11 \\
033459$+$1742.6&       G0V&     71.08&      0.84&    254.71&      0.84& 0.279(4)&12 \\
034814$+$2218.9&       G2V&    112.41&      1.43&    254.38&      2.42& 0.442(7)& 4 \\
034928$+$1254.7&       F8V&    130.48&      1.27&    244.19&      1.28& 0.534(6)&12 \\
035153$-$1031.8&     F3/4V&     65.71&      2.90&    221.10&      9.90& 0.30(02)&11 \\
041209$-$1028.2&       G3V&    107.30&      2.27&    242.57&      2.56&    0.401&11 \\
042925$-$3334.6&     F2/3V&     68.30&     14.90&    229.10&     48.00& 0.30(09)&11 \\
051114$-$0833.4&     F7/8V&    147.98&      2.30&    225.48&      2.35&0.656(12)&13 \\
051832$-$6813.6&     G4/5V&    134.35&      3.46&    213.06&      4.65& 0.63(03)&11 \\
062605$+$2759.9&       F7V&     47.20&      1.16&    279.61&      2.80&0.169(59)& 1 \\
064558$-$0017.5&       G0V&    133.32&      3.06&    198.85&      3.18&0.670(19)&13 \\
071058$-$0352.8&       A8V&    176.07&      0.97&    170.76&      0.97& 1.031(9)& 3 \\
073246$-$2047.5&       F3V&     55.68&      5.52&    226.80&     15.40& 0.25(03)&11 \\
073338$-$5007.4&       F6V&     50.20&     15.40&    265.20&     17.20& 0.19(06)&11 \\
073905$-$0239.1&       F2V&     91.21&      1.83&    244.67&      2.22& 0.373(8)&13 \\
084002$+$1900.0&       F8V&    100.77&      0.82&    221.67&      0.91&0.455(11)& 9 \\
084108$-$3212.1&     A8/9V&    126.22&      2.09&    130.79&      1.52&0.965(20)&11 \\
100141$+$1724.5&       K0V&    144.65&      1.10&    198.41&      1.11& 0.729(7)&10 \\
100234$+$1702.8&    A8/F0V&    262.55&      2.60&     91.49&      1.80&0.348(29)& 2 \\
100248$+$0105.7&     F5/6V&     54.96&      2.30&    281.94&      2.86& 0.195(8)&13 \\
101602$-$0618.5&       F3V&     25.80&      0.45&    258.51&      1.54& 0.100(2)&10 \\
104033$+$1334.0&    A9/F1V&     79.69&      1.12&    262.87&      1.89&0.303(24)& 2 \\
105030$-$0241.7&      F9.5&     74.33&      0.91&    237.30&      1.32& 0.313(5)& 7 \\
110211$+$0953.7&       F5V&    115.56&      0.97&    251.98&      1.17& 0.459(4)&10 \\
110505$+$0509.1&       F8V&     78.20&      2.30&    263.10&      2.90&0.297(09)& 1 \\
120103$+$1300.5&       A5V&     93.39&      1.06&    244.24&      1.97&0.382(21)& 9 \\
121206$+$2232.0&     K4/5V&    124.83&      1.34&    237.00&      1.09& 0.527(6)&10 \\
123300$+$2642.9&     K2/5V&    112.04&      1.28&    237.70&      1.29& 0.471(6)&13 \\
131032$-$0409.5&     K1/2V&    152.78&      0.77&    197.58&      0.77& 0.773(5)&12 \\
134607$+$0506.9&       F8V&    169.39&      1.00&    208.54&      1.00& 0.812(8)& 5 \\
141726$+$1234.1&       G5V&    101.39&      1.07&    236.74&      1.03& 0.428(5)&13 \\
141937$+$0553.8&     F0/1V&    112.68&      0.71&    229.28&      1.24&0.491(11)& 2 \\
143504$+$0906.8&     F7/8V&     31.66&      0.78&    285.31&      1.63&0.111(52)& 2 \\
144803$+$1356.7&       F5V&     64.19&      1.70&    259.07&      1.89& 0.248(7)&13 \\
152243$+$1615.7&    F9/G0V&     40.59&      0.59&    234.24&      0.69&0.173(17)& 3 \\
153152$-$1541.1&       G0V&     84.10&      3.84&    252.20&      4.55&  0.33(4)&5, 15 \\
155649$+$2216.0&       G4V&    138.77&      1.33&    195.64&      1.34& 0.709(8)&13 \\
164121$+$0030.4&       G0V&     82.71&      1.03&    246.70&      1.00& 0.335(9)& 8 \\
165717$+$1059.8&       G5V&     44.12&      1.63&    190.93&      2.90&0.231(10)& 7 \\
171358$+$1621.0&       F4V&     70.52&      1.12&    254.40&      2.27&0.277(24)& 9 \\
173356$+$0810.0&    G0/G1V&    159.64&      0.70&     62.99&      0.62&0.395(12)& 5 \\
175332$-$0354.9&    F8/G2V&     72.06&      2.06&    247.42&      2.08& 0.291(9)&13 \\
180921$+$0909.1&       F7V&     84.82&      1.33&    278.52&      2.01&0.305(24)& 2 \\
182913$+$0647.3&       F7V&     71.33&      0.66&    249.09&      0.89& 0.286(3)&10 \\
185318$+$2113.5&        A7&     64.07&      0.94&    171.91&      0.94& 0.373(6)&12 \\
193524$+$0550.3&       F7V&     97.00&      1.60&    268.20&      2.70& 0.362(7)& 1, 13 \\
194813$+$0918.5&       F9V&    153.03&      0.93&    180.81&      1.14& 0.846(7)&10 \\
203113$+$0513.2&       K2V&    135.74&      1.27&    148.32&      1.27&0.915(12)&13 \\
204628$-$7157.0&    F3IV/V&     52.35&      1.15&    229.34&      3.52& 0.228(8)&14 \\
205710$+$1939.0&     F5/8V&     69.00&      1.60&    184.80&      2.20&0.375(10)& 1, 13 \\
222257$+$1619.4&       F7V&     97.20&      1.50&    270.20&      2.50&0.360(06)& 1, 13  \\
233655$+$1548.1&       F0V&     63.92&      1.12&    250.02&      1.16& 0.256(6)&12 \\
234535$+$2528.3&       F2V&     93.78&      0.86&    234.08&      0.87& 0.401(4)&12 \\
234718$-$0805.2&       F3V&     53.38&      1.53&    263.34&      4.38& 0.203(8)&4  \\
\hline
\end{tabular}}
\end{center}
~~~~~~~1.~\citet{ruci1}; 2.~\citet{ruci2}; 3.~\citet{ruci3}; 4.~\citet{ruci5}; 
5.~\citet{ruci4};  6.~\citet{ruci6}; 7.~\citet{ruci8}; 8.~\citet{ruci9}; 
9.~\citet{ruci11}; 10.~\citet{ruci12}; 11.~\citet{ruci2sp}; 12.~\citet{ruci13}
; 13.~\citet{ruci15}; 14.~\citet{ruci1wi}; 15.~\citet{szal07}
\end{table*}

\section{Determination of Fourier parameters and classification}
\label{sect_clas}
\begin{figure}
\begin{center}
\includegraphics[height=9cm,width=9cm]{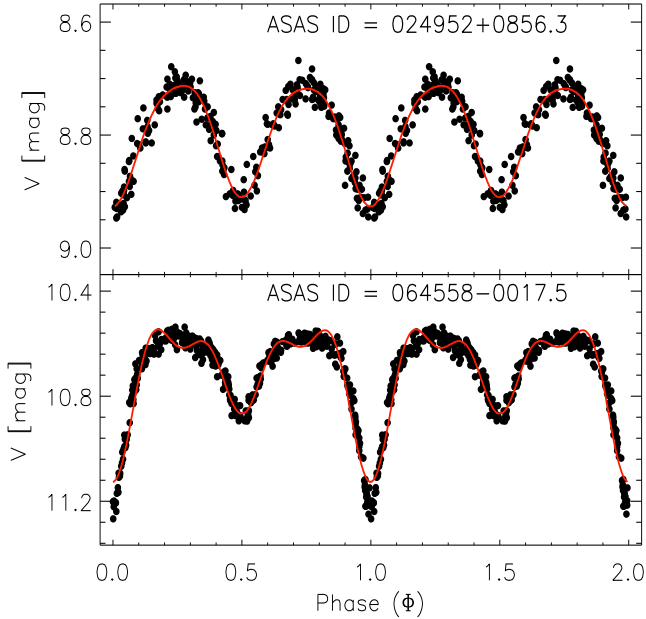}
\end{center}
\caption{Typical Fourier fits for the ASAS eclipsing binaries used for the analysis.}
\label{typical}
\end{figure}

\begin{figure}
\begin{center}
\includegraphics[height=9cm,width=9cm]{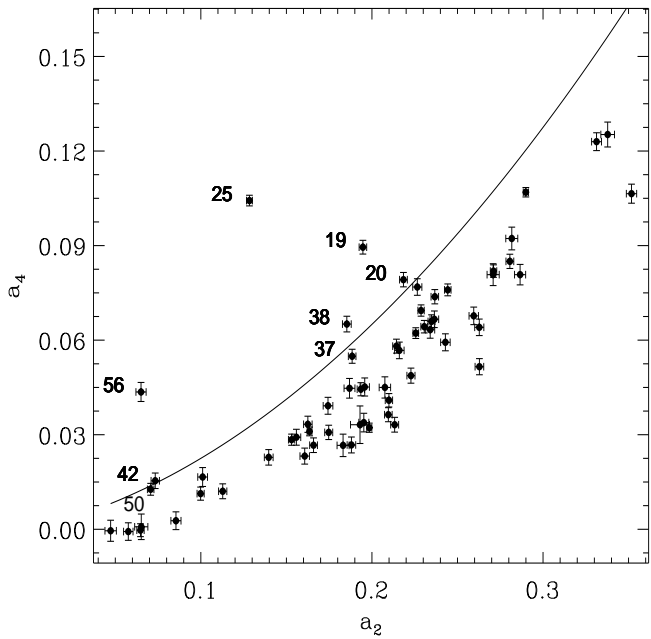}
\end{center}
\caption{Classification of the 62 eclipsing binaries in the a$_{2}-$a$_{4}$ 
plane of Fourier coefficients. Stars lying below the envelope are contact 
binaries and those lying above it are semi-detached and detached binaries. 
The semi-detached and detached binaries in the sample are marked by their 
serial numbers as in Table~\ref{basic_info} corresponding to their respective 
ASAS IDs and described as in the text.}  
\label{class}
\end{figure}

\begin{figure}
\begin{center}
\includegraphics[height=9cm,width=9cm]{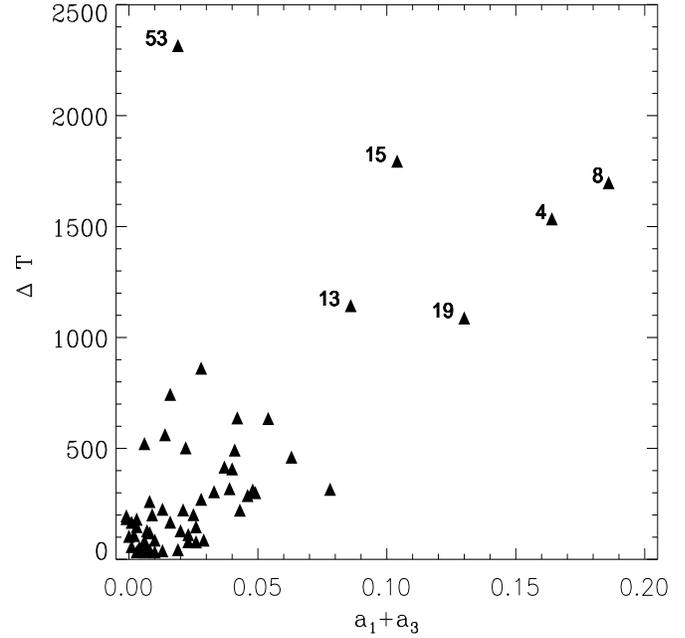}
\end{center}
\caption{Relation between the Fourier coefficients a$_{1}+$a$_{3}$ and the 
temperature difference ($\Delta T \ge 1000$~K) obtained from the modelling for 
the 62 eclipsing binaries in the sample.}
\label{a1a3_temp}
\end{figure}

\citet{ruci93} showed that the light curves of W UMa type systems can be 
quantitatively described by using only the two coefficients a$_{2}$ and 
a$_{4}$ of the cosine decomposition $\sum$ a$_{\rm i} \rm cos(2\pi i\Phi)$. The 
observed light curves were fitted with a Fourier series of the form: 
\begin{equation} 
\rm m(\Phi) = \rm m_{0}+\sum_{i=1}^{4} [a_{i} \rm \, cos(2\pi i \Phi)+b_{i} \rm \, sin(2\pi i \Phi)],
\end{equation}
where $\rm m(\Phi)$ is the phased light curve, $\rm m_{0}$ is the mean 
magnitude and the zero point of the phased light curve corresponds to the 
primary minimum. Examples of the typical Fourier fits to the ASAS light curves 
are shown in Fig.~\ref{typical}. The Fourier parameters of the 62 binaries are
listed in Table~\ref{four}. The resulting distribution of the selected sample
in the a$_{2}$-a$_{4}$ plane is shown in Fig.~\ref{class}. The continuous 
envelope for separating the contact binaries from rest of the sample is shown 
by the relation a$_{4}$ = a$_{2}(0.125+$a$_{2}$). This is similar to the 
envelope relation a$_{4}$ = a$_{2}(0.125-$a$_{2}$) for negative a$_{2}$, 
a$_{4}$ coefficients of \citet{ruci97a}. This gives the envelope for the inner 
contact. The stars lying below this envelope are contact binaries, whereas 
those lying above it are semi-detached and detached binaries. The stars lying 
on or above the envelope can be identified by the serial numbers 19, 20, 25, 
38, 42, 50 and 56 respectively, corresponding to their ASAS IDs 064558$-$0017.5,
 071058$-$0352.8, 084108$-$3212.1, 134607$+$0506.9, 144803$+$1356.7, 
175332$-$0354.9 and 203113$+$0513.2, respectively. Out of these stars, we have 
identified two stars 084108$-$3212.1, 203113$+$0513.2 as detached binaries 
(ED) and the remaining stars as semi-detached binaries (ESD) following the 
boundary lines of \citet{dere07}. In addition, the star with serial number 37 
having ASAS ID 131032$-$0409.5 is found to be of ESD-type from the light curve 
modelling although it lies slightly below the envelope. The stars having ASAS 
IDs 071058$-$0352.8 and 175332$-$0354.9 were also found to be of ESD-types by 
\citet{zola04} and \citet{tase06}, respectively. 
In general, the Fourier coefficients a$_{2}$ relates to the amplitude, 
a$_{1}$ and a$_{3}$ to the difference in the depth of the eclipses and b$_{1}$ 
to the difference between the two maxima \citep{ruci97b,pacz06}. The phenomenon 
of difference in the height of the two maxima is called the \citet{ocon51} 
effect. In Fig.~\ref{a1a3_temp}, we have plotted the temperature 
differences ($\Delta T$) obtained from the  light curve modelling versus 
a$_{1} +$a$_{3}$ for the 62 eclipsing binaries in the 
present sample. The stars with serial numbers 4, 8, 13, 15, 19 and 53 
were found to be binaries having temperature differences of 1000~K or larger 
between the components obtained from the light curve modelling. However, 
sometimes the Fourier fit does not reproduce the bottom of the eclipses very 
well and the algorithm misclassifies the objects. For example, the star having 
serial number 45 with ASAS ID 155649$+$2216.0 may be misclassified as a B-type 
system by the Fourier fitting algorithm because of it gives higher value 
of a$_1$, although, it is not a B-type binary having $\Delta T \sim 316 $K. 
The classification results of eclipsing binary stars based on the single band 
photometry using the Fourier coefficients and without any spectroscopic 
information are generally considered preliminary \citep{pacz06}. It should 
be noted that the star having serial number 53 with ASAS ID 185318$+$2113.5 
is a B-type contact binary obtained from the light curve modelling, although, 
it has smaller values of the Fourier coefficients a$_{1}$ and a$_{3}$. 
We have also found that when the sum (a$_{1} + $ a$_{3}$) is normalised by 
a$_{2}$, the star with serial number 53 moves right on the 
Fig.~\ref{a1a3_temp}. This implies that the coefficients a$_{1}$ and a$_{3}$ 
are obviously inclination dependent and also, to some extent, on amplitude. 
It is obvious that the Fourier method has given smaller values of 
(a$_{1} + $ a$_{3}$) for this very low inclination star. More accurate 
morphological classification of eclipsing binary stars is usually done from 
the photometric light curve modelling with the mass ratio $q$ obtained from 
spectroscopic radial velocity measurements.  
\begin{table*}
\centering
\caption{A sample of the Fourier parameters of the ASAS eclipsing binaries 
used for the analysis. The full table is available as supplementary material 
in the online version of this paper.}
\label{four}
\scalebox{1}{
\begin{tabular}{ccccccccccc}
\\
\hline
\hline
ASAS ID & m$_{0}$& a$_{1}$ & b$_{1}$ & a$_{2}$ & b$_{2}$ & a$_{3}$ &b$_{3}$ & a$_{4}$ & b$_{4}$   \\
&$\sigma$m$_{0}$& $\sigma$a$_{1}$ & $\sigma$b$_{1}$ & $\sigma$a$_{2}$ & $\sigma$b$_{2}$ & $\sigma$a$_{3}$ &$\sigma$b$_{3}$ & $\sigma$a$_{4}$ & $\sigma$b$_{4}$\\\hline
003628$+$2132.3&  11.10009&   0.01911&  -0.00221&   0.18313&  -0.00120&   0.01004&  -0.00030&   0.02662&   0.00128\\
&   0.00248&   0.00344&   0.00359&   0.00352&   0.00356&   0.00344&   0.00365&   0.00356&   0.00348\\
011638$-$3942.5&  10.46070&   0.01523&   0.00084&   0.22563&  -0.00272&   0.00552&   0.00157&   0.06223&  -0.00377\\
&   0.00118&   0.00170&   0.00164&   0.00165&   0.00170&   0.00163&   0.00172&   0.00170&   0.00165\\
012104$+$0736.3&   8.65492&   0.01060&  -0.00076&   0.15583&   0.00097&   0.00168&  -0.00065&   0.02920&  -0.00353\\
&   0.00181&   0.00264&   0.00249&   0.00256&   0.00257&   0.00261&   0.00251&   0.00252&   0.00258\\
014656$-$0945.1&  11.00120&   0.10034&  -0.01179&   0.22871&   0.00806&   0.06391&   0.00053&   0.06940&  -0.00423\\
&   0.00130&   0.00192&   0.00176&   0.00185&   0.00184&   0.00186&   0.00181&   0.00179&   0.00188\\
014854$-$2053.6&  10.53448&   0.01195&  -0.00106&   0.28995&  -0.00044&   0.01128&  -0.00032&   0.10693&   0.00190\\
&   0.00104&   0.00145&   0.00149&   0.00151&   0.00143&   0.00146&   0.00148&   0.00149&   0.00146\\
023833$-$1417.9&   9.63591&   0.00677&  -0.00058&   0.25950&  -0.00234&   0.00605&   0.00021&   0.06771&   0.00220\\
&   0.00195&   0.00273&   0.00280&   0.00274&   0.00279&   0.00281&   0.00272&   0.00278&   0.00275\\
024952$+$0856.3&   8.80081&   0.00902&  -0.00116&   0.10105&   0.00172&  -0.00095&   0.00050&   0.01658&  -0.00235\\
&   0.00205&   0.00299&   0.00283&   0.00285&   0.00297&   0.00297&   0.00285&   0.00300&   0.00283\\
030701$-$5608.1&  11.14192&   0.11353&  -0.00113&   0.23675&   0.00297&   0.07194&   0.00207&   0.07378&   0.00289\\
&   0.00157&   0.00225&   0.00220&   0.00221&   0.00224&   0.00221&   0.00225&   0.00227&   0.00218\\

\hline
\hline
\end{tabular}
}
\end{table*}

\section{Light curve modelling}
\label{sect_mode}
The  mass ratios obtained from radial velocity measurements, mostly 
by RC, were used with the ASAS photometry to derive the geometrical and 
physical parameters of the 62 binary stars. For the light curve modelling, we 
have used the WD code \citep{wil1971,wil1979,wil1990} as implemented in the 
software PHOEBE \citep{prsa05}. In the WD code, some of the parameters need to 
be fixed and the convergence of the solutions were obtained using the methods 
of multiple-subsets. In the code, since most of the binaries were the contact 
binaries, mode-3 version was the most suitable for the light curve analysis. 
However, at the outset, mode 2 (detached) was used to get the convergence of 
the solution and then rapidly ran into mode 3 for the final convergence. In 
the light curve analysis using the WD code, star 1 is the one eclipsed at 
primary minimum and star 2, at the secondary minimum. This notation has been 
used throughout the text.

The input parameters to the WD-code are: the mass ratio $(q)$, 
temperature of the star 1 $T_{1}$, the bolometric albedos $A_{1} = A_{2} = 0.5$
for stars having convective envelopes ($T < 7200$~K) and 1 for radiative 
envelopes ($T > 7200$~K) and gravity darkening coefficients $g_{1} = g_{2} = 0.32$ for convective envelopes and $1.0$ for radiative envelopes, following Von 
Zeipel's law. The mono-chromatic and bolometric limb darkening coefficients 
were interpolated for logarithmic law from van Hamme's table \citep{vanh93}.  

The adjustable parameters are the temperature of the star 2, orbital inclination ($i$), the  surface potentials $\Omega_{1}$, 
$\Omega_{2}$ of the components, monochromatic luminosity of star 1 $L_{1}$. 
Planck function was used to compute the luminosity of the star 1.
In some cases, third light ($l_{3}$) was also kept as an adjustable parameter 
in order to get better fit of some of the light curves. The third light 
($l_{3}$) is in the unit of total light as implemented in PHOEBE. 
However, determination of $l_{3}$ from the photometric light curve 
modelling sometimes affects the orbital inclination, amplitude of the light 
curve. In the two cases (051114$-$0833.4 and 100141$+$1724.5), we have 
found that the determination of third light is highly correlated with the 
orbital inclination. In these cases, we have chosen $l_{3}$ as determined from 
spectroscopy and available in the literature. The values of $l_{3}$ for these 
two stars were 0.16 $\pm 0.02$ \citep{ruci15} and 0.13 \citep{ruci12}, 
respectively. We have adopted these values of $l_{3}$ and kept them fixed in 
the light curve modelling. In the case of two stars 
073246$-$2047.5 and 204628$-$7157.0, we found that their light curves could 
not be fitted properly, especially the minima, using the $q_{sp}$ values. 
Therefore, we have also kept spectroscopic mass ratio $q_{sp}$ as an 
adjustable parameter to get a better fit to these light curve data. 
In these two cases, we have used only the adjusted mass ratios. The values of 
photometric mass ratio $q_{ph}$ obtained for the two cases are $q_{ph} = 
0.190 \pm 0.020, ~0.200 \pm 0.013$, whereas, their spectroscopic values were 
$0.25(3)$ \citep{ruci2sp}, $0.228(8)$ \citep{ruci1wi}, respectively. In the 
case of star 073246$-$2047.5, the $q_{ph}$ value is within 2$\sigma$ of 
$q_{sp}$, whereas for 204628$-$7157.0, the $q_{ph}$ is within 3.5$\sigma$ of 
$q_{sp}$ value, which may be considered a little higher. 
These values of $q_{ph}$ were used to model these two stars. We have marked the 
solutions for these two stars in Table 4. For semi-detached 
binaries, mode 4 and 5 of the WD code were used suitable for the primary and 
the secondary stars filling their limiting lobes respectively. The fill-out 
factor ($f$) is given by 
\begin{eqnarray}
f = \frac{\Omega_{\rm in} - {\Omega_{1,2}}}{\Omega_{\rm in} - \Omega_{\rm out}},
\end{eqnarray}
where $\Omega_{\rm in}$ and $\Omega_{\rm out}$ refer to the inner and the 
outer Lagrangian surface potentials respectively and $\Omega_{1,2}$ 
represents the potential of the surface of star 1, 2 respectively. 
In the case of contact binaries $\Omega_{1} = \Omega_{2} = \Omega$, where 
$\Omega$ is the surface potential of the common envelope for the binary system. 
For semi-detached binaries with the primary component filling its Roche lobe, 
$\Omega_{1} = \Omega_{in}$ and with the secondary filling its Roche lobe 
$\Omega_{2} = \Omega_{in}$. For detached binaries $\Omega_{1,2} > \Omega_{in}$.
Therefore, the eclipsing binaries are classified into contact, semi-detached 
and detached binaries according to $0<f<1$, $f = 0$ and $f < 0$ respectively.
Contact binaries are classified as A-type, if the primary minimum is a 
transit and W-type, if it is an occultation. Therefore $M_{2}/M_{1} < 1$ for 
A-type, $M_{2}/M_{1} > 1$ for W-type contact binaries. The surface 
temperature difference between the two components is in general less than 
1000~K. On the other hand, if the temperature difference between the components 
$\Delta T \ge 1000$~K, then those contact binaries are sometimes called B-type 
\citep{csiz04}. \citet{lucy79} first introduced B-type 
systems which are systems in  geometrical contact, but not in thermal contact,
and therefore, there are large surface temperature  differences between the 
components \citep{csiz04} that produce  differences in the levels of the  
minima. These binaries show an EB ($\beta$ Lyrae)-type light curve, but have 
an orbital period which falls in  the range of classical W UMa systems 
\citep{kalu86}. \citet{ruci97b} defined the B-type contact systems as those 
systems which show large differences in depths of eclipses, yet appear to be 
in contact and exhibit $\beta$ Lyrae-type light curves. Some of them may be 
very close, semi-detached binaries (ESD) mimicking contact systems and some 
may be in contact. 
No attempt has been made to obtain spot parameters for the systems showing 
\citet{ocon51} effect. As it has been observed that with the spot solutions 
included in the WD code, the whole light curve can be easily fitted, but a 
serious problem of uniqueness of the solution persists \citep{mace93} unless 
other means of investigation such as Doppler Imaging techniques are applied 
\citep{mace94}. The stars with serial numbers 15, 34, 43, 48 57 and 60 
exhibit typical \citet{ocon51} effect. A reasonable fit to the maxima of the 
light curves of these stars could be obtained from the light curve modelling
without spots. In this paper, the more massive component is denoted by 
$M_{\rm P}$, and the less massive component by $M_{\rm S}$. Also, throughout 
this paper, unless the errors are written otherwise, we express the standard 
errors in terms of the last quoted digits. For example, 2970.55742(53) should 
be interpreted as 2970.55742 $\pm$ 0.00053.

\begin{figure*}
\begin{center}
\includegraphics[height=14cm,width=18cm]{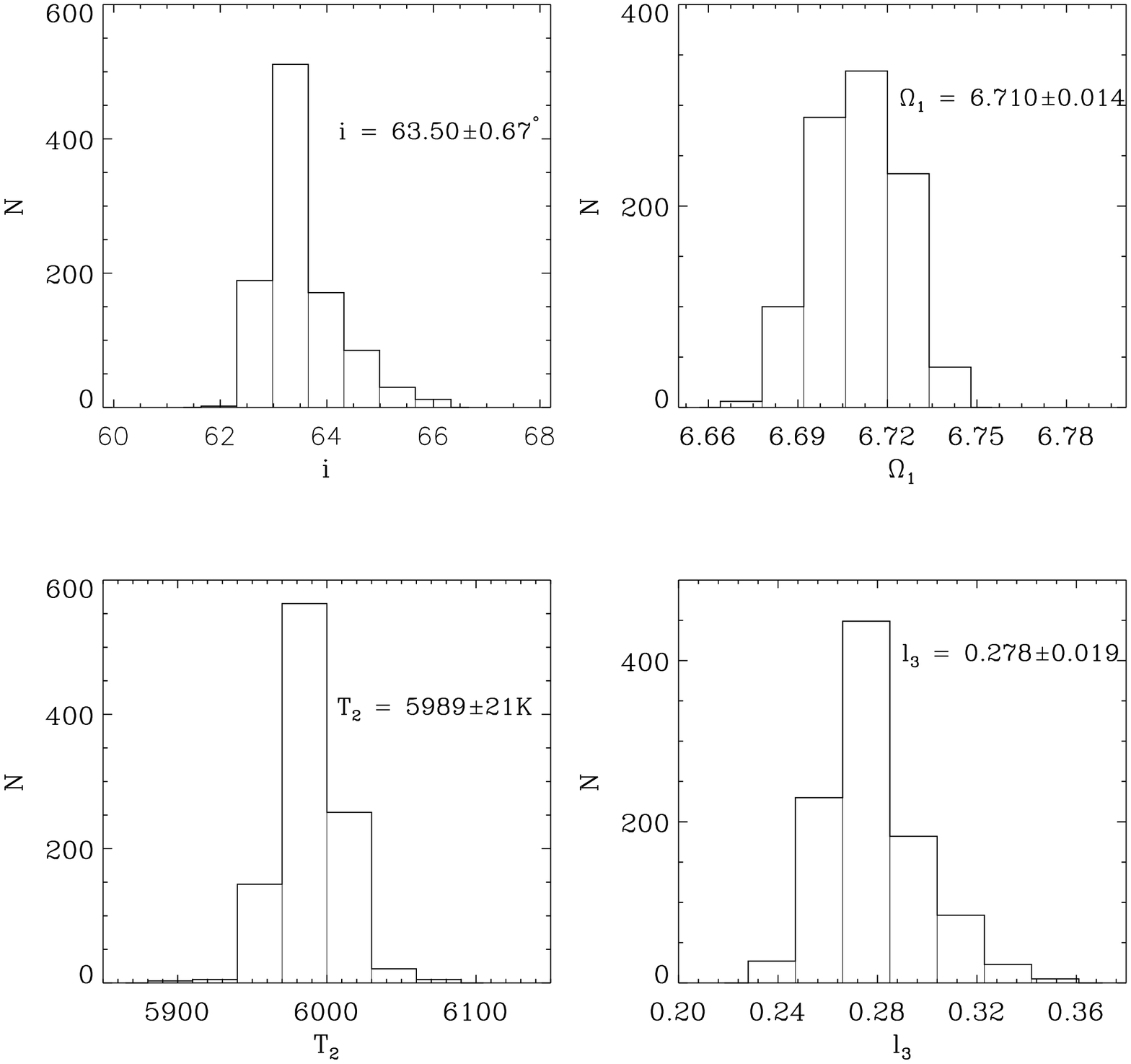}
\end{center}
\caption{Histograms of the results obtained using the Monte Carlo 
parameter scan for the four parameters fit in the overcontact mode with PHOEBE 
for the star ASAS 024952$+$0856.3.}
\label{heuristic}
\end{figure*}

The output of the WD's differential correction ($dc$) minimization program 
as implemented in PHOEBE are the values of the fitting parameters and the 
formal statistical errors associated with each of the fitting parameters. 
Sometimes, the error estimates are too optimistic. The values, errors and 
stability of the solutions were explored using the Monte Carlo parameter scan
(heuristic scan) around the best solution using the PHOEBE's scripter 
capability \citep{bona09}. The WD's $dc$ minimization program was run 1000 
times, each time updating the input parameter values to the values determined 
in the previous iteration. The final values for the parameters were 
determined  from the mean of the parameters obtained in 1000 iterations and 
the standard  deviations were taken as the  errors on these parameters. In 
Fig.~\ref{heuristic}, we show the histograms of the results obtained using the 
heuristic scan method for the four parameters fit in the overcontact mode with 
PHOEBE for the star ASAS 024952$+$0856.3. For the present sample in this 
study, we have found that the error estimates of the fitting parameters 
obtained using the heuristic scan method are reasonable, except for the 
temperature error $\sigma T_{2}$, which is unrealistically small and 
underestimated. In order to obtain a realistic estimate of the error in 
$T_{2}$, we derived the values of $T_{2}$ for two fixed values: $T_{1} - \sigma T_{1}$ and $T_{1} + 
\sigma T_{1} $, keeping other parameters fixed as obtained for $T_{1}$ from 
the modelling, where the value of $\sigma T_{1}$  is obtained from the 
color/spectral type - temperature calibration \citep{cox00} and is listed in 
Table 4. We then find the temperature differences between the newly obtained 
values of $T_{2}$ in the two cases and the $T_{2}$ value obtained for $T_{1}$. 
The mean value of these two temperature differences is taken as the 
approximate error in $T_{2}$. 

\begin{figure*}
\vspace{0.02\linewidth}
\begin{tabular}{cc}
\vspace{+0.01\linewidth}
  \resizebox{0.50\linewidth}{!}{\includegraphics*{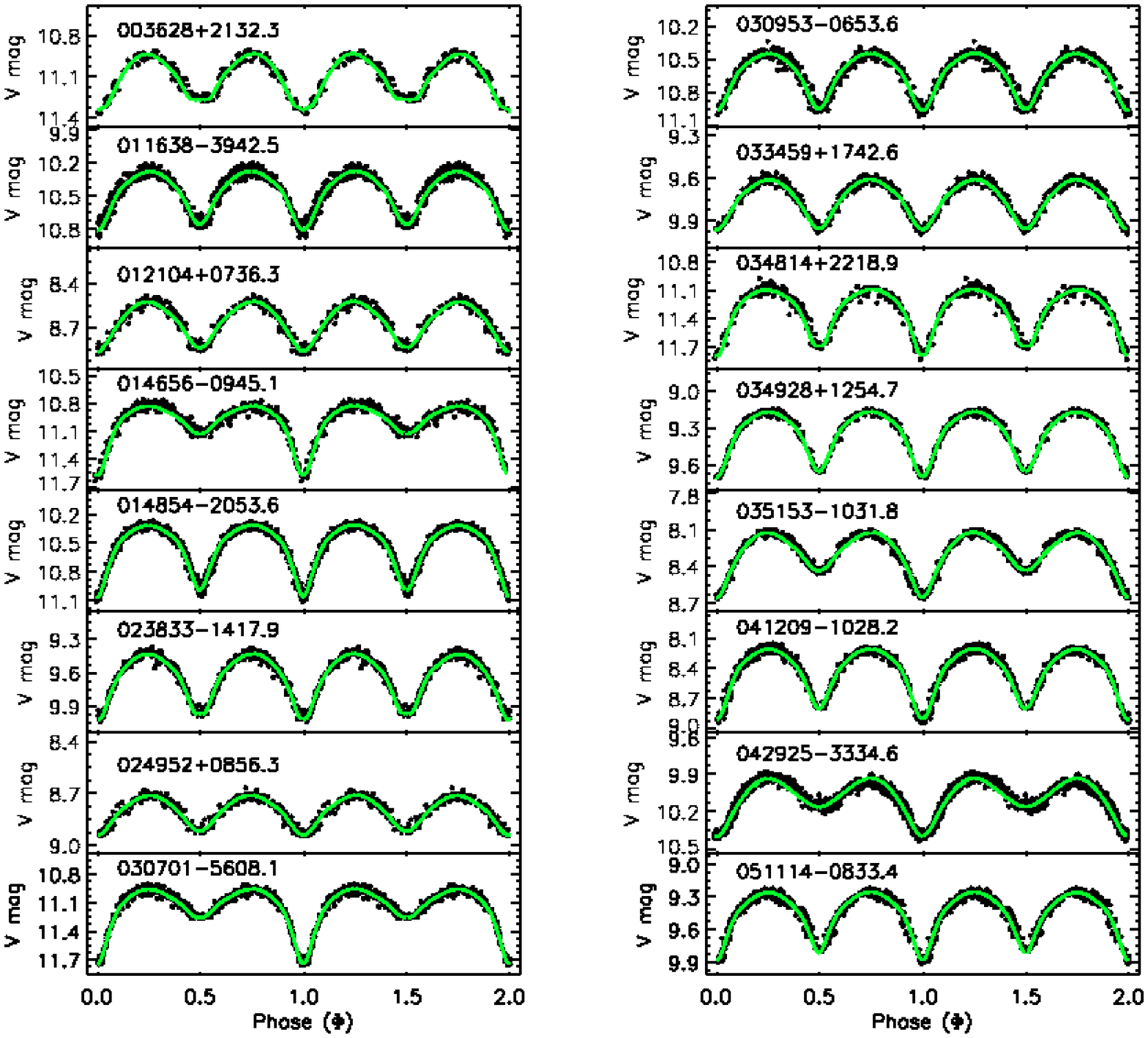}}&
  \resizebox{0.50\linewidth}{!}{\includegraphics*{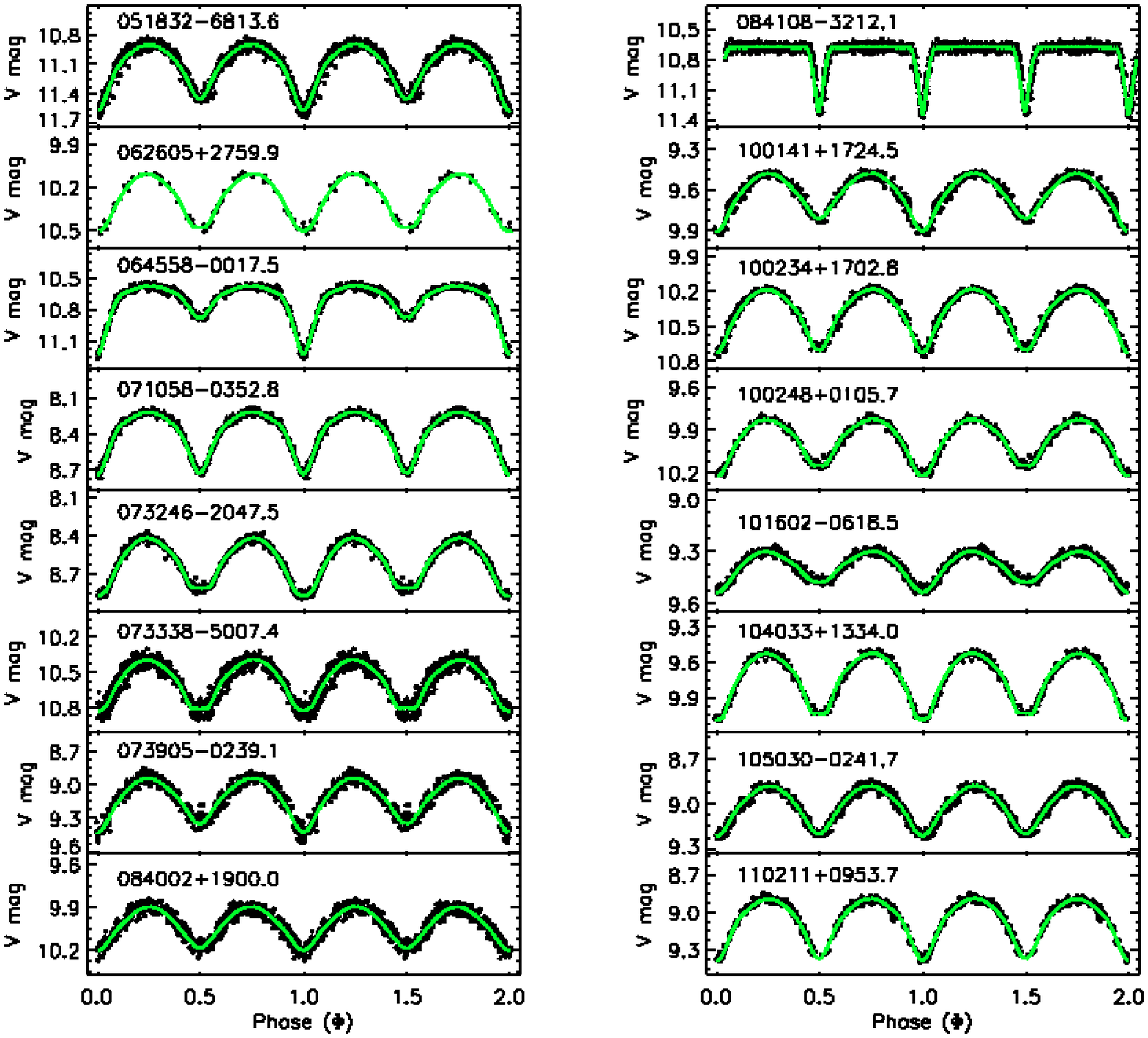}}\\
\vspace{+0.01\linewidth}
  \resizebox{0.50\linewidth}{!}{\includegraphics*{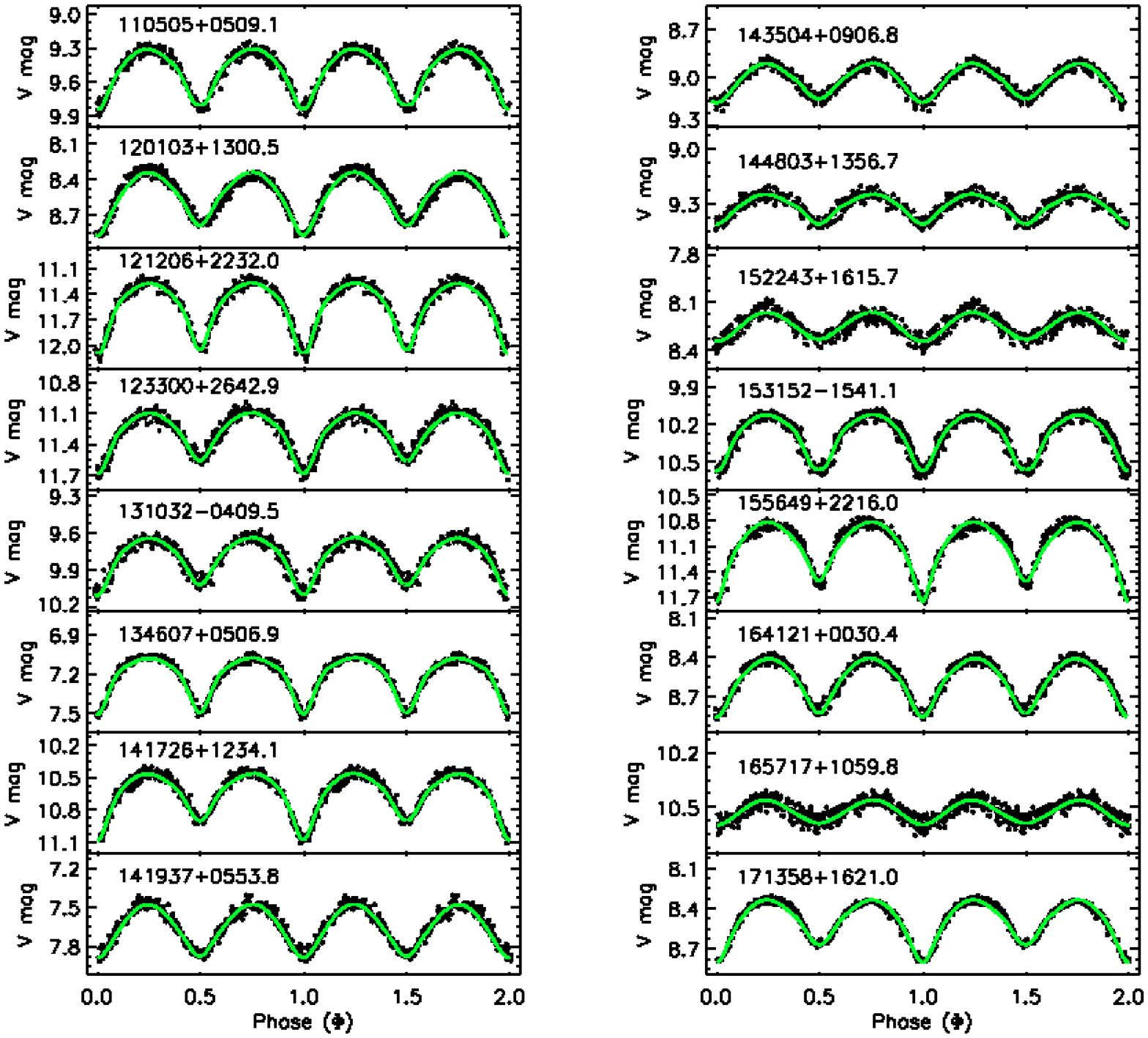}}&
  \resizebox{0.50\linewidth}{!}{\includegraphics*{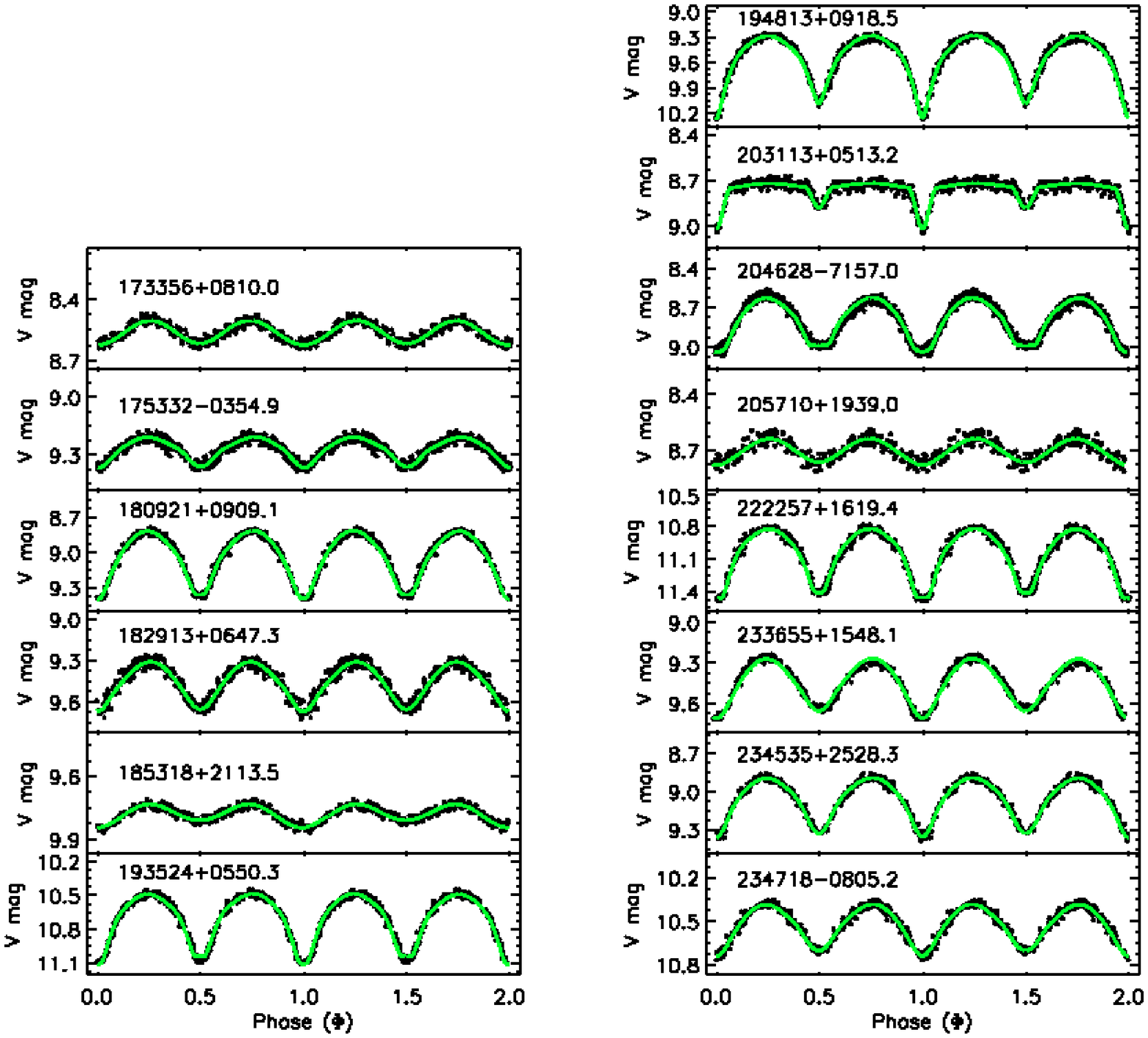}}\\
\vspace{-0.04\linewidth}
\end{tabular}
\caption{Phased light curves of all the 62 stars in the present study. Solid 
line is the synthetic light curve computed from the WD light curve modelling 
technique. The parameters obtained from the modelling are listed in 
Table~\ref{geom1}.}
\label{fig_all}
\end{figure*}

Fig.~\ref{fig_all} shows the phased light curves of all the 62 stars in the 
present study. Solid line is the synthetic light curve computed from the WD 
light curve modelling technique.

\subsection{Temperatures}
To determine the effective temperatures of the primary components of the 
program stars, we used the spectral types of these stars from the literature, 
mostly from RC. Then, theoretical spectral type-temperature relations 
\citep{cox00} were used to calculate the temperatures of the primary 
components. However, when the spectral type is not given in the literature, 
we used the $(J-K)$ color index from 2MASS catalogue \citep{cutr03} to 
determine its spectral type using color-temperature calibration of 
\citet{cox00}. For majority of stars in the sample, the spectral types 
were determined by RC either by direct spectral classification or 
using $(B-V)$ or $(J-K)$ color index. 2MASS data are very useful for spectral 
classification on the basis of the observed colors, especially the $(J-K)$ 
color, with the advantage that the effect of interstellar reddening is 
negligible compared to $(B-V)$ and is monotonically rising from the early 
spectral types to about M0V \citep{cox00, ruci15}. The temperatures of the 
secondary components were estimated by using PHOEBE with the temperatures of 
the primaries fixed. For one star in the sample 110505$+$0509.1 (Ap Leo), we 
have not found any spectral classification. In that case, we estimated its 
spectral type to be F8V corresponding to its infrared color index $(J-K) = 
0.368$ mag. On the other hand, when the spectral classification in RC is given 
as F5/8V, we take its spectral class as F8V.      

\subsection{Absolute dimensions}
In order to calculate the absolute dimensions of the eclipsing binary systems, 
we used the orbital elements obtained from the spectroscopy as given in the 
literature. The orbital elements are the velocity semi-amplitudes 
($K_{1}, K_{2}$). From the light curve modelling, we found  orbital 
inclination $i$, temperatures of the secondary components $T_{2}$, fractional 
stellar radii $r_{1}, r_{2}$. The radii $r_{1,2}$ obtained from the light 
curve modelling are normalised to the semi-major axis of the binary system, 
i.e., $r= R/a$, where $a$ is the semi-major axis and is determined using $i, 
K_{1}, K_{2}, P$. Period $P$ is that obtained from the ME method. The 
component radii were taken as the geometrical mean of the polar, side and back 
radii for contact systems and polar, side, back and point radii for 
semi-detached and detached systems. The luminosities of the two components were
calculated from the radii and temperatures by $L_{j}/L_{\odot} = (R_{j}/R_{\odot})^{2}(T_{j}/5780)^4$, where suffix $j$ refer to either component of the 
system. The mass function of the system is defined as $f(M) = (M_{1} + M_{2})\sin^{3}(i)$, where total mass of the system $M = (M_{1} + M_{2})$ in solar unit 
($M_{\odot}$) was obtained using the Kepler's third law. Hence the individual 
masses were obtained using the total mass and the mass ratio. The spectroscopic
estimate of mass function is given by \citep{ruci2}:
\begin{eqnarray}
f(M) = 1.0385 \times 10^{-7} (K_{1} + K_{2})^3 P ({\rm days})~[M_{\odot}].
\label{mass_func}
\end{eqnarray}
\begin{table*}
\begin{center}
\caption{Results obtained from the light curve modelling}
\label{geom1}
\scalebox{0.88}{
\begin{tabular}{lcccccccccccc}
\hline
ASAS ID & $i[^{o}]$& $\Omega_{1}$ & $\Omega_{2}$ & $x_{1}$ & $x_{2}$& $r_{1}$ &$r_{2}$& $T_{1}[\rm K]^{*} $  & $T_{2}[\rm K]$& fill-out&$l_{3}$  \\ \hline\hline
003628+2132.3&     79.88$\pm$      1.72&     1.985$\pm$     0.010&     1.985&      0.72&      0.72&     0.587$\pm$     0.007&     0.268$\pm$     0.006&6381$\pm$182&6294$  \pm$172&  0.85&		 0\\
011638$-$3942.S&     74.00$\pm$      0.22&     6.096$\pm$     0.008&     6.096&      0.73&      0.73&     0.307$\pm$     0.001&     0.479$\pm$     0.001&5940$\pm$149&5717$\pm$138&  0.19&		 0\\
012104$+$0736.3&     69.06$\pm$      0.80&     2.253$\pm$     0.012&     2.253&      0.73&      0.73&     0.525$\pm$     0.004&     0.274$\pm$     0.005&6250$\pm$157&6024$\pm$150&  0.35&		 0\\
014656$-$0945.1&     82.25$\pm$      0.44&     2.676$\pm$     0.006&     2.676&      0.70&      0.79&     0.464$\pm$     0.001&     0.310$\pm$     0.002&6765$\pm$153&5230$\pm$ 93&  0.08&		 0\\
014854$-$2053.6&     81.18$\pm$      0.10&     3.308$\pm$     0.002&     3.308&      0.73&      0.73&     0.357$\pm$     0.001&     0.408$\pm$     0.001&5865$\pm$152&5753$\pm$147&  0.06&		 0\\
023833$-$1417.9&     82.48$\pm$      0.34&     2.529$\pm$     0.005&     2.529&      0.71&      0.71&     0.486$\pm$     0.001&     0.308$\pm$     0.002&6650$\pm$178&6611$\pm$176&  0.24&		 0\\
024952$+$0856.3&     63.50$\pm$      0.67&     6.710$\pm$     0.014&     6.710&      0.73&      0.73&     0.293$\pm$     0.001&     0.494$\pm$     0.001&6250$\pm$237&5989$\pm$220&  0.02& 0.278$\pm$0.019\\
030701$-$5608.1&     80.36$\pm$      0.17&     2.778$\pm$     0.006&     2.778&      0.70&      0.79&     0.455$\pm$     0.001&     0.323$\pm$     0.001&6881$\pm$169&5183$\pm$ 99&  0.18&		 0\\
030953$-$0653.6&     75.70$\pm$      0.23&     6.065$\pm$     0.008&     6.065&      0.74&      0.74&     0.308$\pm$     0.001&     0.479$\pm$     0.001&6093$\pm$153&6006$\pm$150&  0.18&		 0\\
033459+1742.6&     68.10$\pm$      0.24&     7.251$\pm$     0.011&     7.251&      0.73&      0.73&     0.284$\pm$     0.001&     0.503$\pm$     0.001&5940$\pm$259&5906$\pm$254&  0.18& 0.013$\pm$0.009\\
034814$+$2218.9&     82.33$\pm$      0.46&     2.711$\pm$     0.006&     2.711&      0.73&      0.73&     0.464$\pm$     0.001&     0.322$\pm$     0.002&5790$\pm$128&5711$\pm$119&  0.09&		 0\\
034928$+$1254.7&     74.50$\pm$      0.69&     5.057$\pm$     0.007&     5.057&      0.73&      0.73&     0.326$\pm$     0.001&     0.436$\pm$     0.001&6250$\pm$126&6082$\pm$119&  0.03&		 0\\
035153-1031.8&     74.50$\pm$      0.69&     2.406$\pm$     0.023&     2.406&      0.70&      0.76&     0.504$\pm$     0.007&     0.298$\pm$     0.008&6881$\pm$177&5737$\pm$122&  0.37&		 0\\
041209$-$1028.2&     80.44$\pm$      0.23&     5.547$\pm$     0.008&     5.547&      0.73&      0.73&     0.318$\pm$     0.001&     0.461$\pm$     0.001&5712$\pm$145&5411$\pm$129&  0.16&		 0\\
042925-3334.6&     69.40$\pm$      0.25&     2.337$\pm$     0.004&     2.337&      0.70&      0.80&     0.525$\pm$     0.002&     0.326$\pm$     0.002&6881$\pm$196&5086$\pm$104&  0.71&		 0\\
051114$-$0833.4&     82.08$\pm$      0.44&     4.430$\pm$     0.029&     4.430&      0.72&      0.74&     0.360$\pm$     0.004&     0.433$\pm$     0.004&6250$\pm$153&5931$\pm$134&  0.25&	      0.16 $\pm$ 0.02 $^{a}$\\
051832-6813.6&     76.32$\pm$      0.15&     4.556$\pm$     0.013&     4.556&      0.73&      0.73&     0.351$\pm$     0.002&     0.432$\pm$     0.002&5560$\pm$116&5272$\pm$103&  0.20&		 0\\
062605$+$2759.9&     76.28$\pm$      0.90&     2.081$\pm$     0.007&     2.081&      0.72&      0.72&     0.563$\pm$     0.003&     0.271$\pm$     0.004&6381$\pm$134&6416$\pm$135&  0.66&		 0\\
064558$-$0017.5&     89.00$\pm$      2.60&     3.190&     3.343 $\pm$ 0.052&      0.72&      0.80&     0.442$\pm$     0.002&     0.327$\pm$     0.010&6250$\pm$157&5162$\pm$109&  0.00& 0.242$\pm$0.027\\
071058$-$0352.8&     74.85$\pm$      0.07&     4.037$\pm$     0.011&     3.701&      0.71&      0.72&     0.342$\pm$     0.002&     0.402$\pm$     0.002&7640$\pm$174&7744$\pm$182&  0.00&		 0\\
073246$-$2047.5&     84.47$\pm$      0.58&     2.121$\pm$     0.002&     2.121&      0.70&      0.70&     0.557$\pm$     0.001&     0.284$\pm$     0.001&6881$\pm$315&6801$\pm$304&  0.71&		 0 $^{\dag}$\\
073338$-$5007.4&     79.47$\pm$      0.73&     2.134$\pm$     0.004&     2.134&      0.71&      0.71&     0.553$\pm$     0.002&     0.278$\pm$     0.002&6514$\pm$144&6662$\pm$148&  0.57&		 0\\
073905$-$0239.1&     72.14$\pm$      0.27&     2.512$\pm$     0.005&     2.512&      0.70&      0.70&     0.496$\pm$     0.002&     0.329$\pm$     0.002&7000$\pm$191&6584$\pm$169&  0.49&		 0\\
084002$+$1900.0&     62.19$\pm$      0.48&     5.427$\pm$     0.021&     5.427&      0.73&      0.73&     0.323$\pm$     0.003&     0.460$\pm$     0.002&6250$\pm$127&6121$\pm$120&  0.23&		 0\\
084108$-$3212.1&     90.10$\pm$      0.97&     6.004$\pm$     0.056&6.640 $\pm$0.095&0.67&      0.67&     0.200$\pm$     0.002&     0.172$\pm$     0.002&7468$\pm$203&7521$\pm$208&  $<0$&		 0\\
100141$+$1724.5&     68.63$\pm$      0.37&     4.214$\pm$     0.022&     4.214&      0.75&      0.75&     0.368$\pm$     0.003&     0.423$\pm$     0.003&5150$\pm$118&4742$\pm$248&  0.25&	      0.13 $^{b}$\\
100234$+$1702.8&     77.16$\pm$      0.42&     2.471$\pm$     0.010&     2.471&      0.69&      0.69&     0.500$\pm$     0.003&     0.321$\pm$     0.004&7300$\pm$170&7265$\pm$167&  0.40&		 0\\
100248$+$0105.7&     79.12$\pm$      0.44&     2.146$\pm$     0.009&     2.146&      0.70&      0.72&     0.550$\pm$     0.003&     0.279$\pm$     0.005&6650$\pm$163&6448$\pm$153&  0.52& 0.114$\pm$0.011\\
101602$-$0618.5&     74.88$\pm$      0.65&     1.928$\pm$     0.002&     1.928&      0.70&      0.72&     0.594$\pm$     0.001&     0.221$\pm$     0.001&6881$\pm$199&6378$\pm$169&  0.42&		 0\\
104033$+$1334.0&     84.05$\pm$      0.89&     2.374$\pm$     0.004&     2.374&      0.69&      0.69&     0.514$\pm$     0.002&     0.313$\pm$     0.002&7148$\pm$163&7158$\pm$167&  0.83&		 0\\
105030$-$0241.7&     65.53$\pm$      0.18&     6.720$\pm$     0.007&     6.720&      0.73&      0.73&     0.294$\pm$     0.001&     0.493$\pm$     0.001&5960$\pm$182&5877$\pm$176&  0.20&		 0\\
110211$+$0953.7&     73.85$\pm$      0.15&     5.400$\pm$     0.006&     5.400&      0.70&      0.70&     0.324$\pm$     0.001&     0.460$\pm$     0.001&6650$\pm$148&6616$\pm$144&  0.18&		 0\\
110505$+$0509.1&     79.66$\pm$      0.65&     2.404$\pm$     0.012&     2.404&      0.73&      0.73&     0.503$\pm$     0.004&     0.295$\pm$     0.004&6250$\pm$155&6247$\pm$156&  0.37&		 0\\
120103$+$1300.5&     72.54$\pm$      0.30&     2.547$\pm$     0.006&     2.547&      0.70&      0.70&     0.489$\pm$     0.002&     0.324$\pm$     0.002&8180$\pm$188&7542$\pm$167&  0.40&		 0\\
121206$+$2232.0&     90.58$\pm$      2.85&     4.982$\pm$     0.027&     4.982&      0.79&      0.79&     0.338$\pm$     0.004&     0.450$\pm$     0.003&4546$\pm$ 72&4399$\pm$ 67&  0.04&		 0\\
123300$+$2642.9&     72.43$\pm$      0.29&     5.319$\pm$     0.009&     5.319&      0.77&      0.77&     0.326$\pm$     0.001&     0.457$\pm$     0.001&4830$\pm$115&4517$\pm$ 98&  0.15&		  0\\
131032$-$0409.5&     68.91$\pm$      0.22&     4.231$\pm$     0.019&     4.207&      0.77&      0.77&     0.382$\pm$     0.003&     0.428$\pm$     0.001&4830$\pm$136&4559$\pm$121&  0.00&		 0\\
134607$+$0506.9&     79.06$\pm$      0.43&     3.437&     4.998 $\pm$ 0.077&      0.73&      0.73&     0.356$\pm$     0.002&     0.360$\pm$     0.001&6250$\pm$145&6211$\pm$146&  0.00&		 0\\
141726$+$1234.1&     77.29$\pm$      0.39&     2.698$\pm$     0.014&     2.698&      0.73&      0.73&     0.463$\pm$     0.003&     0.316$\pm$     0.004&5560$\pm$120&5099$\pm$101&  0.12&		 0\\
141937$+$0553.8&     65.52$\pm$      0.43&     2.696$\pm$     0.017&     2.696&      0.69&      0.69&     0.482$\pm$     0.005&     0.362$\pm$     0.006&7148$\pm$198&7201$\pm$203&  0.60&		 0\\
143504$+$0906.8&     60.87$\pm$      0.89&     1.924$\pm$     0.004&     1.924&      0.73&      0.73&     0.601$\pm$     0.001&     0.255$\pm$     0.004&6250$\pm$219&6234$\pm$216&  0.91&		 0\\
144803$+$1356.7&     73.96$\pm$      1.00&     2.348&     2.978 $\pm$ 0.086&      0.70&      0.70&     0.535$\pm$     0.002&     0.157$\pm$     0.009&6650$\pm$175&6846$\pm$187&  0.00& 0.302$\pm$0.036\\
152243$+$1615.7&     50.47$\pm$      2.24&     2.090$\pm$     0.017&     2.090&      0.73&      0.73&     0.552$\pm$     0.005&     0.260$\pm$     0.006&5940$\pm$144&5759$\pm$283&  0.68&		 0\\
153152$-$1541.1&     90.35$\pm$      2.40&     2.479$\pm$     0.007&     2.479&      0.73&      0.73&     0.492$\pm$     0.002&     0.302$\pm$     0.002&5790$\pm$162&5974$\pm$166&  0.29& 0.207$\pm$0.008\\
155649$+$2216.0&     83.85$\pm$      0.67&     3.157$\pm$     0.058&     3.157&      0.73&      0.73&     0.446$\pm$     0.007&     0.349$\pm$     0.007&5636$\pm$108&5320$\pm$ 92&  0.46&		 0\\
164121$+$0030.4&     69.77$\pm$      0.21&     6.397$\pm$     0.013&     6.397&      0.73&      0.73&     0.307$\pm$     0.002&     0.495$\pm$     0.001&5940$\pm$136&5739$\pm$129&  0.22&		 0\\
165717$+$1059.8&     47.05$\pm$      0.68&     2.261$\pm$     0.007&     2.261&      0.73&      0.73&     0.525$\pm$     0.002&     0.267$\pm$     0.001&5560$\pm$135&5440$\pm$127&  0.21&		 0\\
171358+1621.0&     88.69$\pm$      5.29&     2.374$\pm$     0.029&     2.374&      0.70&      0.74&     0.506$\pm$     0.009&     0.286$\pm$     0.001&6765$\pm$168&6130$\pm$139&  0.30&		 0\\
173356$+$0810.0&     44.06$\pm$      1.24&     5.954$\pm$     0.027&     5.954&      0.73&      0.73&     0.303$\pm$     0.003&     0.465$\pm$     0.003&5940$\pm$165&5418$\pm$332&  0.14&		 0\\
175332$-$0354.9&     57.97$\pm$      0.31&     2.505$\pm$     0.007&     2.446&      0.73&      0.73&     0.491$\pm$     0.003&     0.299$\pm$     0.001&5940$\pm$151&5773$\pm$144&  0.00& 0.144$\pm$0.008\\
180921$+$0909.1&     80.19$\pm$      0.25&     2.363$\pm$     0.003&     2.363&      0.71&      0.71&     0.518$\pm$     0.001&     0.320$\pm$     0.001&6381$\pm$135&6424$\pm$136&  0.61&		 0\\
182913$+$0647.3&     66.36$\pm$      0.48&     6.991$\pm$     0.023&     6.991&      0.72&      0.72&     0.301$\pm$     0.003&     0.514$\pm$     0.002&6381$\pm$191&6273$\pm$180&  0.56&		 0\\
185318$+$2113.5&     41.59$\pm$      2.78&     2.569$\pm$     0.034&     2.569&      0.72&      0.80&     0.481$\pm$     0.009&     0.311$\pm$     0.010&7816$\pm$173&5500$\pm$623&  0.41&		 0\\
193524$+$0550.3&     82.47$\pm$      0.22&     6.160$\pm$     0.005&     6.160&      0.72&      0.72&     0.307$\pm$     0.001&     0.483$\pm$     0.001&6381$\pm$153&6252$\pm$147&  0.24&		 0\\
194813$+$0918.5&     93.81$\pm$      0.22&     3.384$\pm$     0.006&     3.384&      0.74&      0.75&     0.413$\pm$     0.001&     0.384$\pm$     0.001&6093$\pm$152&5871$\pm$140&  0.24&		 0\\
203113$+$0513.2&     77.06$\pm$      1.76&     6.017$\pm$     0.591&4.384 $\pm$0.094&0.77&      0.77&     0.200$\pm$     0.021&     0.281$\pm$     0.008&4830$\pm$ 94&4337$\pm$ 68&  $<0$&		 0\\
204628$-$7157.0&     84.81$\pm$      0.60&     2.147$\pm$     0.003&     2.147&      0.70&      0.70&     0.552$\pm$     0.001&     0.287$\pm$     0.002&6881$\pm$160&6837$\pm$158&  0.52& 0.089$\pm$0.005 $^{\dag}$\\
205710$+$1939.0&     45.25$\pm$      2.45&     6.051$\pm$     0.080&     6.051&      0.73&      0.75&     0.308$\pm$     0.009&     0.478$\pm$     0.008&6250$\pm$185&6192$\pm$168&  0.09&		 0\\
222257$+$1619.4&     87.96$\pm$      2.59&     6.142$\pm$     0.012&     6.142&      0.72&      0.74&     0.311$\pm$     0.001&     0.487$\pm$     0.002&6381$\pm$126&6076$\pm$113&  0.32&		 0\\
233655$+$1548.1&     71.11$\pm$      0.67&     2.240$\pm$     0.009&     2.240&      0.70&      0.71&     0.470$\pm$     0.003&     0.312$\pm$     0.003&7300$\pm$184&6438$\pm$141&  0.81&		 0\\
234535$+$2528.3&     73.23$\pm$      0.33&     2.640$\pm$     0.012&     2.640&      0.69&      0.69&     0.541$\pm$     0.003&     0.319$\pm$     0.006&7000$\pm$192&6438$\pm$161&  0.10&		 0\\
234718$-$0805.2&     70.36$\pm$      0.32&     2.177$\pm$     0.004&     2.177&      0.70&      0.70&     0.542$\pm$     0.002&     0.276$\pm$     0.002&6881$\pm$182&6137$\pm$140&  0.48&		 0\\
\hline

\end{tabular}
}
\end{center}
$^{*}$The errors in $T_{1}$ are obtained from the color/spectral type-temperature calibration \citep{cox00} and are not obtained from the modelling. 
${\dag}$The solutions are obtained using the $q_{ph}$ values as described in 
the text. $^{a}$\citet{ruci15}, $^{b}$\citet{ruci12}.
\end{table*}

\begin{table*}
\begin{center}
\caption{ Physical parameters of the 62 eclipsing binary stars}
\label{phys}
\scalebox{0.88}{
\begin{tabular}{cccccccccc}
\hline
ID & $a$ [in R$_{\odot}$]&$M/M_{\odot}$ & $f(M)$[in $M_{\odot}$]&  $M_{1}$/$M_{\odot}$& $M_{2}$/$M_{\odot}$ &$R_{1}$/$R_{\odot}$&$R_{2}$/$R_{\odot}$ 
&$L_{1}$/$L_{\odot}$&$L_{2}$/$L_{\odot}$  \\\hline
003628$+$2132.3&     2.490$\pm$     0.039&     1.542$\pm$     0.070&     1.471$\pm$     0.071&     1.358$\pm$     0.055&     0.185$\pm$     0.020&     1.462$\pm$     0.029&     0.667$\pm$     0.018&     3.174$\pm$     0.487&     0.626$\pm$     0.103\\
011638$-$3942.5&     2.592$\pm$     0.027&     1.616$\pm$     0.049&     1.435$\pm$     0.044&     0.436$\pm$     0.041&     1.179$\pm$     0.151&     0.796$\pm$     0.009&     1.242$\pm$     0.013&     0.706$\pm$     0.086&     1.476$\pm$     0.174\\
012104$+$0736.3&     3.253$\pm$     0.026&     2.037$\pm$     0.048&     1.659$\pm$     0.047&     1.661$\pm$     0.026&     0.375$\pm$     0.010&     1.708$\pm$     0.019&     0.891$\pm$     0.018&     3.987$\pm$     0.489&     0.937$\pm$     0.131\\
014656$-$0945.1&     3.726$\pm$     0.280&     2.932$\pm$     0.644&     2.852$\pm$     0.627&     2.109$\pm$     0.264&     0.823$\pm$     0.215&     1.729$\pm$     0.130&     1.155$\pm$     0.087&     5.608$\pm$     1.352&     0.894$\pm$     0.199\\
014854$-$2053.6&     2.329$\pm$     0.032&     1.685$\pm$     0.067&     1.625$\pm$     0.065&     0.722$\pm$     0.053&     0.963$\pm$     0.097&     0.831$\pm$     0.012&     0.950$\pm$     0.013&     0.733$\pm$     0.097&     0.886$\pm$     0.115\\
023833$-$1417.9&     3.046$\pm$     0.027&     1.948$\pm$     0.051&     1.898$\pm$     0.050&     1.436$\pm$     0.034&     0.511$\pm$     0.025&     1.480$\pm$     0.014&     0.938$\pm$     0.010&     3.840$\pm$     0.482&     1.507$\pm$     0.194\\
024952$+$0856.3&     2.948$\pm$     0.024&     2.377$\pm$     0.056&     1.704$\pm$     0.050&     0.569$\pm$     0.029&     1.808$\pm$     0.126&     0.864$\pm$     0.008&     1.456$\pm$     0.012&     1.020$\pm$     0.173&     2.445$\pm$     0.400\\
030701$-$5608.1&     4.277$\pm$     0.189&     2.680$\pm$     0.345&     2.568$\pm$     0.330&     1.823$\pm$     0.159&     0.857$\pm$     0.147&     1.946$\pm$     0.086&     1.381$\pm$     0.061&     7.605$\pm$     1.419&     1.234$\pm$     0.203\\
030953$-$0653.6&     3.069$\pm$     0.022&     1.952$\pm$     0.041&     1.776$\pm$     0.038&     0.530$\pm$     0.027&     1.421$\pm$     0.098&     0.945$\pm$     0.008&     1.470$\pm$     0.011&     1.103$\pm$     0.128&     2.519$\pm$     0.290\\
033459$+$1742.6&     2.779$\pm$     0.011&     1.796$\pm$     0.021&     1.435$\pm$     0.018&     0.392$\pm$     0.015&     1.404$\pm$     0.070&     0.789$\pm$     0.004&     1.398$\pm$     0.006&     0.695$\pm$     0.129&     2.130$\pm$     0.385\\
034814$+$2218.9&     2.500$\pm$     0.019&     1.795$\pm$     0.041&     1.747$\pm$     0.040&     1.245$\pm$     0.028&     0.550$\pm$     0.021&     1.160$\pm$     0.009&     0.805$\pm$     0.008&     1.355$\pm$     0.142&     0.618$\pm$     0.064\\
034928$+$1254.7&     2.349$\pm$     0.014&     1.862$\pm$     0.032&     1.666$\pm$     0.033&     0.648$\pm$     0.021&     1.214$\pm$     0.054&     0.766$\pm$     0.005&     1.024$\pm$     0.006&     0.802$\pm$     0.075&     1.286$\pm$     0.117\\
035153$-$1031.8&     2.990$\pm$     0.108&     1.388$\pm$     0.146&     1.242$\pm$     0.132&     1.068$\pm$     0.123&     0.320$\pm$     0.057&     1.507$\pm$     0.058&     0.891$\pm$     0.040&     4.561$\pm$     0.822&     0.770$\pm$     0.135\\
041209$-$1028.2&     2.257$\pm$     0.022&     1.489$\pm$     0.043&     1.428$\pm$     0.041&     0.426$\pm$     0.036&     1.063$\pm$     0.115&     0.718$\pm$     0.007&     1.041$\pm$     0.010&     0.491$\pm$     0.060&     0.832$\pm$     0.096\\
042925$-$3334.6&     3.987$\pm$     0.674&     2.110$\pm$     1.041&     1.730$\pm$     0.854&     1.623$\pm$     0.577&     0.487$\pm$     0.320&     2.093$\pm$     0.354&     1.300$\pm$     0.220&     8.801$\pm$     3.978&     1.013$\pm$     0.425\\
051114$-$0833.4&     3.159$\pm$     0.028&     2.354$\pm$     0.061&     2.288$\pm$     0.060&     0.933$\pm$     0.034&     1.422$\pm$     0.078&     1.137$\pm$     0.016&     1.368$\pm$     0.018&     1.768$\pm$     0.223&     2.074$\pm$     0.241\\
051832$-$6813.6&     2.020$\pm$     0.034&     1.354$\pm$     0.066&     1.242$\pm$     0.061&     0.523$\pm$     0.063&     0.830$\pm$     0.128&     0.709$\pm$     0.013&     0.873$\pm$     0.015&     0.430$\pm$     0.051&     0.527$\pm$     0.059\\
062605$+$2759.9&     3.289$\pm$     0.033&     1.951$\pm$     0.057&     1.789$\pm$     0.056&     1.669$\pm$     0.034&     0.282$\pm$     0.013&     1.852$\pm$     0.021&     0.891$\pm$     0.016&     5.094$\pm$     0.544&     1.206$\pm$     0.145\\
064558$-$0017.5&     3.734$\pm$     0.050&     2.160$\pm$     0.084&     2.159$\pm$     0.084&     1.294$\pm$     0.051&     0.867$\pm$     0.059&     1.650$\pm$     0.023&     1.221$\pm$     0.041&     3.724$\pm$     0.479&     0.949$\pm$     0.143\\
071058$-$0352.8&     4.814$\pm$     0.019&     3.258$\pm$     0.038&     2.930$\pm$     0.034&     1.654$\pm$     0.016&     1.604$\pm$     0.028&     1.646$\pm$     0.012&     1.935$\pm$     0.012&     8.274$\pm$     0.871&    12.067$\pm$     1.288\\
073246$-$2047.5&     4.396$\pm$     0.266&     1.694$\pm$     0.300&     1.671$\pm$     0.296&     1.422$\pm$     0.205&     0.272$\pm$     0.079&     2.448$\pm$     0.148&     1.248$\pm$     0.076&   12.041$\pm$     3.665&     2.987$\pm$     0.897\\
073338$-$5007.4&     2.746$\pm$     0.201&     1.482$\pm$     0.317&     1.408$\pm$     0.301&     1.245$\pm$     0.270&     0.237$\pm$     0.125&     1.519$\pm$     0.111&     0.763$\pm$     0.056&     3.721$\pm$     0.875&     1.029$\pm$     0.243\\
073905$-$0239.1&     4.453$\pm$     0.039&     2.907$\pm$     0.074&     2.507$\pm$     0.065&     2.117$\pm$     0.032&     0.790$\pm$     0.029&     2.209$\pm$     0.021&     1.465$\pm$     0.016&    10.494$\pm$     1.347&     3.614$\pm$     0.448\\
084002$+$1900.0&     2.762$\pm$     0.016&     1.924$\pm$     0.033&     1.331$\pm$     0.029&     0.602$\pm$     0.020&     1.322$\pm$     0.057&     0.892$\pm$     0.010&     1.271$\pm$     0.009&     1.088$\pm$     0.112&     2.030$\pm$     0.189\\
084108$-$3212.1&    11.791$\pm$     0.119&     4.083$\pm$     0.120&     4.083$\pm$     0.120&     2.078$\pm$     0.040&     2.005$\pm$     0.079&     2.358$\pm$     0.033&     2.028$\pm$     0.031&    15.499$\pm$     2.125&    11.792$\pm$     1.667\\
100141$+$1724.5&     2.071$\pm$     0.011&     1.473$\pm$     0.022&     1.190$\pm$     0.020&     0.621$\pm$     0.020&     0.852$\pm$     0.035&     0.762$\pm$     0.007&     0.876$\pm$     0.008&     0.366$\pm$     0.041&     0.348$\pm$     0.079\\
100234$+$1702.8&     3.505$\pm$     0.032&     2.422$\pm$     0.064&     2.245$\pm$     0.061&     1.797$\pm$     0.033&     0.625$\pm$     0.025&     1.752$\pm$     0.019&     1.125$\pm$     0.017&     7.813$\pm$     0.898&     3.159$\pm$     0.388\\
100248$+$0105.7&     2.850$\pm$     0.031&     1.758$\pm$     0.056&     1.665$\pm$     0.054&     1.471$\pm$     0.040&     0.287$\pm$     0.020&     1.568$\pm$     0.019&     0.795$\pm$     0.017&     4.305$\pm$     0.528&     0.979$\pm$     0.134\\
101602$-$0618.5&     3.148$\pm$     0.021&     1.431$\pm$     0.028&     1.288$\pm$     0.028&     1.301$\pm$     0.022&     0.130$\pm$     0.006&     1.870$\pm$     0.013&     0.696$\pm$     0.006&     7.021$\pm$     0.909&     0.717$\pm$     0.088\\
104033$+$1334.0&     4.212$\pm$     0.028&     2.619$\pm$     0.051&     2.577$\pm$     0.051&     2.010$\pm$     0.024&     0.609$\pm$     0.017&     2.165$\pm$     0.017&     1.318$\pm$     0.012&    10.964$\pm$     1.168&     4.089$\pm$     0.457\\
105030$-$0241.7&     3.004$\pm$     0.016&     1.846$\pm$     0.029&     1.392$\pm$     0.022&     0.440$\pm$     0.019&     1.406$\pm$     0.080&     0.883$\pm$     0.006&     1.481$\pm$     0.008&     0.882$\pm$     0.119&     2.345$\pm$     0.308\\
110211$+$0953.7&     2.770$\pm$     0.012&     2.126$\pm$     0.026&     1.884$\pm$     0.024&     0.669$\pm$     0.016&     1.457$\pm$     0.048&     0.897$\pm$     0.005&     1.274$\pm$     0.006&     1.411$\pm$     0.140&     2.787$\pm$     0.269\\
110505$+$0509.1&     2.954$\pm$     0.033&     1.864$\pm$     0.060&     1.775$\pm$     0.058&     1.437$\pm$     0.040&     0.427$\pm$     0.025&     1.486$\pm$     0.020&     0.872$\pm$     0.015&     3.019$\pm$     0.382&     1.037$\pm$     0.140\\
120103$+$1300.5&     4.501$\pm$     0.031&     2.956$\pm$     0.059&     2.566$\pm$     0.053&     2.139$\pm$     0.024&     0.817$\pm$     0.021&     2.201$\pm$     0.018&     1.458$\pm$     0.013&    19.432$\pm$     2.096&     6.165$\pm$     0.660\\
121206$+$2232.0&     1.580$\pm$     0.008&     1.085$\pm$     0.015&     1.084$\pm$     0.015&     0.374$\pm$     0.018&     0.710$\pm$     0.043&     0.534$\pm$     0.007&     0.711$\pm$     0.006&     0.109$\pm$     0.010&     0.170$\pm$     0.013\\
123300$+$2642.9&     1.723$\pm$     0.009&     1.216$\pm$     0.019&     1.053$\pm$     0.017&     0.389$\pm$     0.020&     0.826$\pm$     0.054&     0.562$\pm$     0.004&     0.787$\pm$     0.005&     0.154$\pm$     0.017&     0.231$\pm$     0.023\\
131032$-$0409.5&     2.313$\pm$     0.008&     1.710$\pm$     0.017&     1.389$\pm$     0.015&     0.745$\pm$     0.013&     0.964$\pm$     0.023&     0.883$\pm$     0.008&     0.990$\pm$     0.004&     0.381$\pm$     0.049&     0.379$\pm$     0.043\\
134607$+$0506.9&     3.105$\pm$     0.012&     2.412$\pm$     0.028&     2.283$\pm$     0.029&     1.331$\pm$     0.015&     1.081$\pm$     0.021&     1.105$\pm$     0.008&     1.118$\pm$     0.005&     1.671$\pm$     0.178&     1.666$\pm$     0.173\\
141726$+$1234.1&     2.348$\pm$     0.011&     1.479$\pm$     0.020&     1.373$\pm$     0.020&     1.036$\pm$     0.017&     0.443$\pm$     0.012&     1.087$\pm$     0.009&     0.742$\pm$     0.010&     1.012$\pm$     0.103&     0.333$\pm$     0.035\\
141937$+$0553.8&     3.574$\pm$     0.019&     2.645$\pm$     0.042&     1.994$\pm$     0.037&     1.774$\pm$     0.019&     0.871$\pm$     0.017&     1.723$\pm$     0.020&     1.294$\pm$     0.023&     6.941$\pm$     0.931&     4.033$\pm$     0.595\\
143504$+$0906.8&     2.550$\pm$     0.026&     1.760$\pm$     0.053&     1.173$\pm$     0.047&     1.584$\pm$     0.034&     0.176$\pm$     0.008&     1.533$\pm$     0.016&     0.650$\pm$     0.012&     3.211$\pm$     0.518&     0.572$\pm$     0.101\\
144803$+$1356.7&     2.754$\pm$     0.026&     1.633$\pm$     0.044&     1.450$\pm$     0.045&     1.309$\pm$     0.033&     0.325$\pm$     0.017&     1.473$\pm$     0.015&     0.432$\pm$     0.025&     3.804$\pm$     0.477&     0.368$\pm$     0.083\\
152243$+$1615.7&     2.093$\pm$     0.068&     1.393$\pm$     0.132&     0.639$\pm$     0.087&     1.187$\pm$     0.099&     0.205$\pm$     0.020&     1.155$\pm$     0.039&     0.544$\pm$     0.022&     1.488$\pm$     0.245&     0.292$\pm$     0.081\\
153152$-$1541.1&     2.384$\pm$     0.042&     1.414$\pm$     0.073&     1.413$\pm$     0.073&     1.063$\pm$     0.065&     0.351$\pm$     0.039&     1.173$\pm$     0.021&     0.720$\pm$     0.014&     1.385$\pm$     0.205&     0.592$\pm$     0.088\\
155649$+$2216.0&     2.572$\pm$     0.015&     1.526$\pm$     0.026&     1.499$\pm$     0.026&     0.893$\pm$     0.022&     0.633$\pm$     0.023&     1.147$\pm$     0.019&     0.898$\pm$     0.019&     1.190$\pm$     0.131&     0.578$\pm$     0.064\\
164121$+$0030.4&     3.150$\pm$     0.014&     2.035$\pm$     0.027&     1.681$\pm$     0.023&     0.511$\pm$     0.017&     1.524$\pm$     0.071&     0.967$\pm$     0.008&     1.559$\pm$     0.008&     1.043$\pm$     0.112&     2.362$\pm$     0.236\\
165717$+$1059.8&     2.641$\pm$     0.047&     1.428$\pm$     0.075&     0.560$\pm$     0.035&     1.160$\pm$     0.061&     0.268$\pm$     0.025&     1.386$\pm$     0.025&     0.705$\pm$     0.013&     1.646$\pm$     0.220&     0.390$\pm$     0.051\\
171358$+$1621.0&     2.711$\pm$     0.022&     1.501$\pm$     0.035&     1.500$\pm$     0.037&     1.175$\pm$     0.028&     0.326$\pm$     0.014&     1.372$\pm$     0.027&     0.775$\pm$     0.007&     3.531$\pm$     0.489&     0.760$\pm$     0.082\\
173356$+$0810.0&     2.695$\pm$     0.061&     1.448$\pm$     0.096&     0.487$\pm$     0.046&     0.410$\pm$     0.071&     1.038$\pm$     0.192&     0.817$\pm$     0.020&     1.253$\pm$     0.030&     0.744$\pm$     0.120&     1.212$\pm$     0.355\\
175332$-$0354.9&     3.181$\pm$     0.031&     2.368$\pm$     0.067&     1.443$\pm$     0.044&     1.834$\pm$     0.036&     0.534$\pm$     0.026&     1.562$\pm$     0.018&     0.951$\pm$     0.010&     2.720$\pm$     0.339&     0.900$\pm$     0.108\\
180921$+$0909.1&     2.984$\pm$     0.020&     2.127$\pm$     0.041&     2.035$\pm$     0.040&     1.630$\pm$     0.024&     0.497$\pm$     0.016&     1.546$\pm$     0.011&     0.955$\pm$     0.007&     3.550$\pm$     0.350&     1.392$\pm$     0.138\\
182913$+$0647.3&     2.598$\pm$     0.013&     1.666$\pm$     0.025&     1.281$\pm$     0.024&     0.370$\pm$     0.017&     1.295$\pm$     0.073&     0.782$\pm$     0.009&     1.335$\pm$     0.009&     0.908$\pm$     0.129&     2.473$\pm$     0.315\\
185318$+$2113.5&     3.470$\pm$     0.191&     2.299$\pm$     0.369&     0.672$\pm$     0.154&     1.675$\pm$     0.169&     0.625$\pm$     0.073&     1.669$\pm$     0.097&     1.079$\pm$     0.069&     9.317$\pm$     1.907&     0.955$\pm$     0.554\\
193524$+$0550.3&     2.699$\pm$     0.023&     1.920$\pm$     0.048&     1.871$\pm$     0.047&     0.510$\pm$     0.031&     1.410$\pm$     0.113&     0.829$\pm$     0.008&     1.304$\pm$     0.012&     1.020$\pm$     0.117&     2.327$\pm$     0.260\\
194813$+$0918.5&     3.355$\pm$     0.015&     1.969$\pm$     0.025&     1.956$\pm$     0.025&     1.067$\pm$     0.017&     0.902$\pm$     0.023&     1.386$\pm$     0.007&     1.288$\pm$     0.007&     2.371$\pm$     0.260&     1.767$\pm$     0.187\\
203113$+$0513.2&     3.009$\pm$     0.029&     1.340$\pm$     0.037&     1.240$\pm$     0.043&     0.700$\pm$     0.034&     0.640$\pm$     0.040&     0.602$\pm$     0.063&     0.845$\pm$     0.025&     0.177$\pm$     0.051&     0.227$\pm$     0.028\\
204628$-$7157.0&     4.450$\pm$     0.059&     1.866$\pm$     0.072&     1.843$\pm$     0.071&     1.520$\pm$     0.045&     0.346$\pm$     0.019&     2.456$\pm$     0.033&     1.277$\pm$     0.019&    12.118$\pm$     1.449&     3.193$\pm$     0.390\\
205710$+$1939.0&     2.573$\pm$     0.112&     1.723$\pm$     0.220&     0.617$\pm$     0.111&     0.470$\pm$     0.138&     1.253$\pm$     0.401&     0.792$\pm$     0.042&     1.230$\pm$     0.058&     0.933$\pm$     0.206&     1.992$\pm$     0.403\\
222257$+$1619.4&     2.630$\pm$     0.021&     1.863$\pm$     0.044&     1.860$\pm$     0.045&     0.493$\pm$     0.029&     1.370$\pm$     0.105&     0.818$\pm$     0.007&     1.281$\pm$     0.012&     0.915$\pm$     0.090&     2.003$\pm$     0.185\\
233655$+$1548.1&     4.182$\pm$     0.027&     2.413$\pm$     0.046&     2.044$\pm$     0.046&     1.922$\pm$     0.023&     0.492$\pm$     0.015&     1.965$\pm$     0.018&     1.305$\pm$     0.015&     9.828$\pm$     1.170&     2.620$\pm$     0.290\\
234535$+$2528.3&     3.920$\pm$     0.016&     2.409$\pm$     0.029&     2.115$\pm$     0.028&     1.720$\pm$     0.015&     0.690$\pm$     0.013&     2.120$\pm$     0.015&     1.250$\pm$     0.024&     9.673$\pm$     1.195&     2.406$\pm$     0.333\\
234718$-$0805.2&     3.254$\pm$     0.065&     1.990$\pm$     0.117&     1.663$\pm$     0.098&     1.655$\pm$     0.070&     0.336$\pm$     0.034&     1.764$\pm$     0.036&     0.898$\pm$     0.019&     6.248$\pm$     0.917&     1.025$\pm$     0.137\\

\hline
	 
\end{tabular}
}
\end{center}
\end{table*}

\section{Results and comparison with the literature} 
\label{sect_resu}
In this section, we present determinations of the geometrical and physical 
parameters of the 62 eclipsing binary stars using the WD code and the orbital 
elements as given in the literature. Table~\ref{geom1} lists the parameters 
obtained from the light curve modelling. In Table~\ref{phys}, we list the 
physical parameters of all the 62 stars determined using the parameters in 
Table~\ref{geom1} and the radial velocity semi-amplitudes as listed in 
Table~\ref{radial_info}.
\subsection{Comparison with 203113$+$0513.2 = MR Del}
In order to check the quality of the results obtained, we compare the spectral 
type of the system MR Del (203113$+$0513.2) as given in \citet{ruci15} to that 
determined by us from the light curve modelling. \citet{ruci15} predicted a 
spectral classification (K2V+K6V) for the system. Spectral type K2V is 
consistent with the infrared color index of $(J-K) = 0.620$ mag. Assuming this 
as the spectral type of the primary component, we calculated its temperature 
as $T_{1} = 4830$ K following the calibration of \citet{cox00}. From the light 
curve modelling, we found the temperature of the secondary component as 
$T_{2} = 4337 \pm 68$~K, which is consistent with the spectral type K6V for 
the secondary component as estimated by \citet{ruci15}.

\begin{figure*}
\begin{center}
\includegraphics[width=17cm,height=23cm]{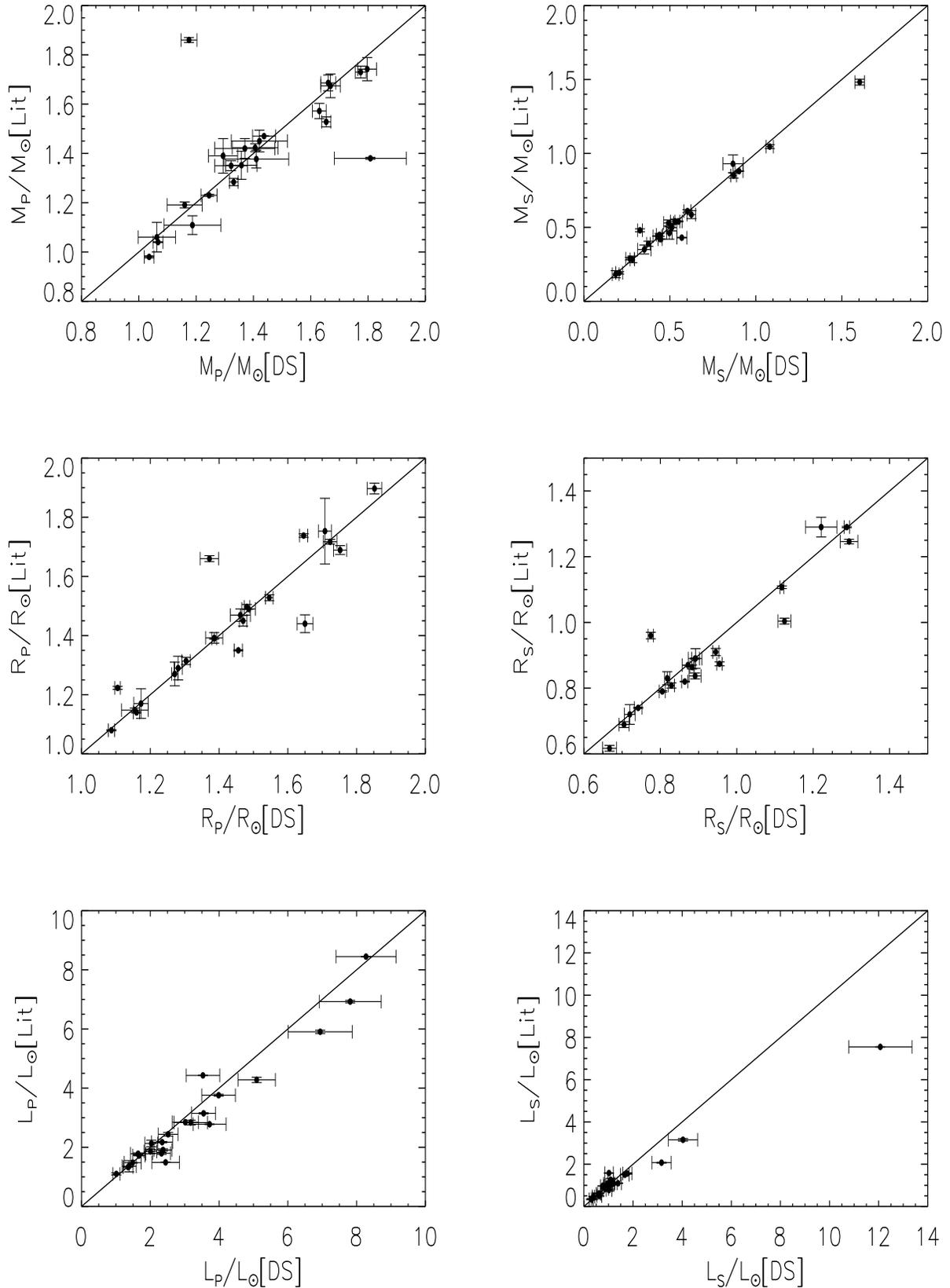}
\end{center}
\vspace{-0.01\linewidth}
\caption{The comparison of primary and secondary component parameters of 23 
stars in our sample common to literature. The parameters determined by us are 
plotted on the X-axis while those in the literature, on the Y-axis. The 
clustering of most of the data points along a line (solid) indicates a 
positive linear correlation.}
\label{comparison}
\end{figure*}

We also compare the determination of physical parameters of 23 stars common to 
our sample with the previous recent publications. We list all the 23 stars in 
Table~\ref{table_common} and their physical parameters along with the 
references in the literature. In Fig.~\ref{comparison}, we show 
scatter plots between the parameters determined by us (shown on the X-axis and 
represented by `DS' in the parenthesis) and those in the literature (shown on 
the Y-axis and represented by `Lit' in the parenthesis). From 
Fig.~\ref{comparison}, it seems that the errors in $L_{1}$ and $L_{2}$ were 
underestimated in some of the cases in the literature and their calculations 
were not clearly mentioned. The clustering of data points along a straight 
line of zero intercept and unit slope indicates a positive linear correlation 
with the presence of three outliers. The ASAS IDs of these stars are 
024952$+$0856.3 (EE Cet), 071058$-$0352.8 (V753 Mon), and 171358$+$1621.0 
(AK Her). We discuss these stars in the following subsections in view of 
mismatch in their parameter determinations.
\begin{table*}
\begin{center}
\caption{ Physical parameters of 23 stars common to our sample as found in the 
literature}
\label{table_common}
\scalebox{0.88}{
\begin{tabular}{cccccccl}
\hline
ID & $M_{\rm P}$/$M_{\odot}$& $M_{\rm S}$/$M_{\odot}$ &$R_{\rm P}$/$R_{\odot}$&$R_{\rm S}$/$R_{\odot}$ &$L_{\rm P}$/$L_{\odot}$&$L_{\rm S}$/$L_{\odot}$&References  \\\hline
003628$+$2132.3&     1.352$\pm$     0.057&     0.183$\pm$     0.024&     1.469$\pm$     0.021&     0.617$\pm$     0.009&     2.836$\pm$     0.081&     0.493$\pm$     0.022&\citet{gaze05} \\
012104$+$0736.3&     1.682$\pm$     0.032&     0.389$\pm$     0.017&     1.753$\pm$     0.111&     0.890$\pm$     0.002&     3.760$\pm$     0.032&     0.984$\pm$     0.008&\citet{gaze06}\\
024952$+$0856.3&     1.380$\pm$     0.004&     0.430$\pm$     0.002&     1.350$\pm$     0.002&     0.820$\pm$     0.001&     1.490$\pm$     0.007&     1.580$\pm$     0.009&\citet{djur06}\\
030953$-$0653.6&     1.450$\pm$     0.044&     0.540$\pm$     0.017&     1.450$\pm$     0.018&     0.910$\pm$     0.011&     2.440$\pm$     0.061&     1.030$\pm$     0.025&\citet{qian07b}\\
034814$+$2218.9&     1.230$\pm$     0.004&     0.540$\pm$     0.002&     1.140$\pm$     0.002&     0.790$\pm$     0.002&     1.330$\pm$     0.004&     0.610$\pm$     0.002&\citet{yuan07}\\
062605$+$2759.9&     1.674$\pm$     0.048&     0.283$\pm$     0.022&     1.897$\pm$     0.018&     0.837$\pm$     0.009&     4.279$\pm$     0.090&     1.090$\pm$     0.058&\citet{gaze05}\\
064558$-$0017.5&     1.390$\pm$     0.070&     0.930$\pm$     0.060&     1.440$\pm$     0.030&     1.290$\pm$     0.030&     2.780$\pm$     0.011&     1.080$\pm$     0.005&\citet{gaze09}\\
071058$-$0352.8&     1.528$\pm$     0.020&     1.482$\pm$     0.020&     1.738$\pm$     0.006&     1.592$\pm$     0.007&     8.446$\pm$     0.011&     7.551$\pm$     0.012&\citet{zola04}\\
084002$+$1900.0&     1.350$\pm$     0.020&     0.610$\pm$     0.010&     1.270$\pm$     0.040&     0.890$\pm$     0.030&     2.130$\pm$     0.110&     1.260$\pm$     0.070&\citet{zhan09}\\
100234$+$1702.8&     1.742$\pm$     0.047&     0.586$\pm$     0.027&     1.689$\pm$     0.015&     1.004$\pm$     0.009&     6.926$\pm$     0.043&     2.073$\pm$     0.020&\citet{gaze06}\\
105030$-$0241.7&     1.423$\pm$     0.016&     0.449$\pm$     0.009&     1.497$\pm$     0.008&     0.864$\pm$     0.006&     2.174$\pm$     0.022&     0.832$\pm$     0.008&\citet{gaze06}\\
110505$+$0509.1&     1.470$\pm$     0.000&     0.440$\pm$     0.000&     1.490$\pm$     0.000&     0.870$\pm$     0.000&     2.846$\pm$     0.000&     1.003$\pm$     0.000&\citet{qian07a}\\
134607$+$0506.9&     1.284$\pm$     0.015&     1.046$\pm$     0.013&     1.223$\pm$     0.005&     1.107$\pm$     0.004&     1.720$\pm$     0.002&     1.500$\pm$     0.001&\citet{zola05}\\
141726$+$1234.1&     0.980$\pm$     0.004&     0.420$\pm$     0.002&     1.080$\pm$     0.002&     0.740$\pm$     0.002&     1.100$\pm$     0.036&     0.360$\pm$     0.006&\citet{rain90}\\
141937$+$0553.8&     1.730$\pm$     0.024&     0.850$\pm$     0.017&     1.717$\pm$     0.008&     1.246$\pm$     0.007&     5.905$\pm$     0.055&     3.155$\pm$     0.060&\citet{gaze05}\\
152243$+$1615.7&     1.109$\pm$     0.038&     0.192$\pm$     0.008&     1.148$\pm$     0.007&     0.507$\pm$     0.006&     1.480$\pm$     0.010&     0.340$\pm$     0.020&\citet{zola05}\\
153152$-$1541.1&     1.060$\pm$     0.060&     0.350$\pm$     0.030&     1.170$\pm$     0.050&     0.720$\pm$     0.030&     1.360$\pm$     0.190&     0.590$\pm$     0.050&\citet{szal07}\\
165717$+$1059.8&     1.191$\pm$     0.012&     0.288$\pm$     0.009&     1.392$\pm$     0.018&     0.689$\pm$     0.009&     1.782$\pm$     0.030&     0.468$\pm$     0.011&\citet{gaze06}\\
171358$+$1621.0&     1.860$\pm$     0.010&     0.480$\pm$     0.010&     1.660$\pm$     0.010&     0.960$\pm$     0.010&     4.430$\pm$     0.010&     0.970$\pm$     0.010&\citet{sama10}\\
180921$+$0909.1&     1.572$\pm$     0.031&     0.462$\pm$     0.042&     1.528$\pm$     0.010&     0.874$\pm$     0.006&     3.148$\pm$     0.020&     1.097$\pm$     0.010&\citet{gaze06}\\
193524$+$0550.3&     1.377$\pm$     0.036&     0.498$\pm$     0.022&     1.314$\pm$     0.012&     0.808$\pm$     0.008&     1.796$\pm$     0.033&     0.777$\pm$     0.041&\citet{gaze06}\\
194813$+$0918.5&     1.040$\pm$     0.002&     0.880$\pm$     0.002&     1.390$\pm$     0.002&     1.290$\pm$     0.002&     1.910$\pm$     0.039&     1.560$\pm$     0.032&\citet{hriv89}\\
222257$+$1619.4&     1.420$\pm$     0.040&     0.530$\pm$     0.020&     1.290$\pm$     0.040&     0.830$\pm$     0.020&     1.860$\pm$     0.080&     0.940$\pm$     0.060&\citet{kalo07}\\

\hline
	 
\end{tabular}
}
\end{center}
\end{table*}
\subsection{ASAS 024952+0856.3 = EE Cet}
Variability of the combined light of the visual binary ADS 2163 was discovered 
by the Hipparcos satellite mission \citep{hipp97}. The system shows a 
light-curve variation with an amplitude of 0.23 mag and a 0.38 day period in 
the Hipparcos photometric observations. \citet{ruci6} correctly found that the 
source of variability is the southern, fainter companion, which is classified 
as a W UMa type eclipsing binary EE Cet (P $\sim$ 0.38 days). In their radial 
velocity study, EE Cet was observed without contamination by the northern 
companion, separated by 5.6$^{''}$. They assigned a spectral class F8V to the 
system and classified it as a W-type W UMa binary with a mass ratio  
$q_{sp} = 0.315(5)$. A photometric study of this star was carried out by 
\citet{djur06} using $q = 0.315$ who obtained the  parameters as listed in 
Table~\ref{table_eecet}. They obtained $i[^{o}] = 78.5 \pm 0.2$, 
$M_{1} + M_{2} = 1.80~M_{\odot}, L_{3} \sim 0.54 $ and mass function 
$f(M) = 1.694~M_{\odot}$. This value of $f(M) = 1.694~M_{\odot}$ is quite 
close to the spectroscopically determined value $1.706(41)~M_{\odot}$ of 
\citet{ruci6}. We have also used the same value of $q_{sp} = 0.315$ to model 
the ASAS light curve. Since the star is a W-type contact binary, we have used 
the inverse of $q_{sp}$ in the photometric light curve modelling technique and 
obtained $i[^{o}] = 63.50 \pm 0.67, f(M) = 1.704(50)~M_{\odot}, l_{3} = 0.278
\pm 0.019$. The mass function $f(M)$ in all the three cases is quite similar. 
However, the absolute parameters determined using the ASAS light curve are 
different from those of \citet{djur06}. This disagreement might be caused 
because of slightly noisy ASAS data and/or estimation of different values of 
third light contribution in the two analyses. Simultaneous multi-color 
photometry and the use of spectral broadening function using high resolution 
radial velocity curves can resolve the issue of accurate estimation of third 
light in the system. This will help in the determination of reliable parameters for this binary system.            

\subsection{ASAS 071058-0352.8 = V753 Mon}
V753 Mon is a new discovery of the Hipparcos mission. The radial velocity 
measurements of the star were obtained by \citet{ruci3} who found a
spectroscopic mass ratio of $q_{sp} = 0.970 $. The star was classified as a 
W UMa binary of subtype-W with spectral type A8V by the authors. The recent
photometric analysis of the star by \citet{zola04} yields the following 
parameters of the binary: $i[^{o}] = 75.30 \pm 0.13, \Delta T \sim 120$~K, 
using a semi-detached configuration with the slightly less massive component 
filling its Roche lobe. Our photometric analysis of the ASAS light curve also 
shows that the configurations of the system is not a contact one. The star is 
marked with the serial number `20' in Fig.~\ref{class} and lies above the 
contact envelope. We have also found a semi-detached configuration with the 
secondary star filling its Roche lobe from the ASAS light curve modelling.  
We found : $i[^{o}] = 74.85 \pm 0.07, \Delta T \sim 104 $~K. We now 
compare our mass function determination with \citet{zola04} and \citet{ruci3}.
The physical parameters obtained by \citet{zola04} and those determined by us 
are listed in Table~\ref{table_v753mon}. The mass function estimated from 
\citet{zola04} is $2.724~M_{\odot}$ following equation~(\ref{mass_func}). 
On the other hand, the mass function obtained from the ASAS light curve is  
$f(M) = 2.930(34)~M_{\odot}$ and is identical to the spectroscopic estimate 
$2.93(6)~M_{\odot}$ of \citet{ruci3}. However, the mass function derived from 
\citet{zola04} is quite different from both the spectroscopic estimate of 
\citet{ruci3} and that determined by us. Therefore the parameters determined 
by \citet{zola04} need to re-examined. 
\begin{table}
\begin{center}
\caption{Parameters for the components of EE Cet and comparison 
with previous study.}
\label{table_eecet}
\begin{tabular}{ccl}
\hline\hline
Parameter & \citet{djur06} & This study \\\hline
$q[\rm fixed]$ &0.315 &0.315  \\
$M_{\rm P}($\rm M$_{\odot}$)&1.37(4) &1.808(126) \\
$M_{\rm S}($\rm M$_{\odot}$)&0.43(2) &0.569(29) \\
$R_{\rm P}($\rm R$_{\odot}$)&1.35(2) &1.456(12)\\
$R_{\rm S}($\rm R$_{\odot}$)&0.82(1) &0.864(8)\\
$L_{\rm P}($\rm L$_{\odot}$)&1.58(9) &2.445(400) \\
$L_{\rm S}($\rm L$_{\odot}$)&1.49(7) &1.020(173) \\
\hline
\end{tabular}
\end{center}
\end{table}

\begin{table}
\begin{center}
\caption{Parameters for the components of V753 Mon and comparison 
with previous study.}
\label{table_v753mon}
\begin{tabular}{ccl}
\hline\hline
Parameter & \citet{zola04}&This study \\\hline
$q[\rm fixed]$ &0.971 &0.970  \\
$M_{\rm P}($M$_{\odot}$)&1.528(20) &1.654(16) \\
$M_{\rm S}($M$_{\odot}$)&1.482(20) &1.604(28) \\
$R_{\rm P}($R$_{\odot}$)&1.738(7)  &1.646(12)\\
$R_{\rm S}($R$_{\odot}$)&1.592(6)  &1.935(12)\\
$L_{\rm P}($L$_{\odot}$)&8.446(68) &8.274(871)\\
$L_{\rm S}($L$_{\odot}$)&7.551(112)&12.067(1.288)\\
\hline
\end{tabular}
\end{center}
\end{table}

\subsection{ASAS 171358+1621.0 = AK Her}
AK Her is the brighter component in the visual binary 
ADS 10408. The companion, located at a separation of 4.2$^{''}$, is 3.5 mag 
fainter than AK Her at its maximum light \citep{ruci11}. The star has been the 
subject of numerous studies \citep[][and references therein]{sama10}. These 
suggest that the system has a variable light curve and a variable period. 
Analysis of the updated $O-C$ curves revealed evidence of continuously 
changing period with cycle length between 57 and 97 years \citep{sama10}. 
Using the spectroscopic mass ratio of 0.26 from \citet{sanf34}, \citet{sama10} 
derived $i[^{o}] = 83.88 \pm 0.27$ and the parameters as listed in 
Table~\ref{table_akher}. Using the spectroscopic radial velocity measurements, 
\citet{ruci11} obtained $K_{1} = 70.52 \pm 1.12$ km~s$^{-1}$, $K_{2} = 254.40 
\pm 2.27$ km~s$^{-1}$. \citet{ruci11} quoted $f(M) = 1.598(29)$, while the 
correct value should be 1.502(29), which can be calculated from 
equation~(\ref{mass_func}). From \citet{sama10}, one can calculate the value 
of mass function $f(M) = (M_{1}+M_{2})\sin^{3}(i) = 2.300~M_{\odot}$. This 
value is quite different from the spectroscopic estimate $1.502(29)$. Hence, 
the parameters determined by \citet{sama10} may not be reliable. We have 
analysed the ASAS light curve of AK Her using the spectroscopic mass ratio 
$q_{sp} = 0.277$, considering A-type configuration and a spectral type F4V 
\citep{ruci11}. Using the ASAS data, we calculated the mass function for 
the system as $1.500(37) M_{\odot}$, which is nearly identical to the 
spectroscopic estimate of \citet{ruci11}. 
\begin{table}
\begin{center}
\caption{Parameters for the components of AK her and comparison 
with previous study.}
\label{table_akher}
\begin{tabular}{ccl}
\hline\hline
Parameter &\citet{sama10} &This study \\\hline
$q[\rm fixed]$&0.26&0.277  \\
$M_{\rm P}($M$_{\odot}$)&1.86(1) &1.175(28) \\
$M_{\rm S}($M$_{\odot}$)&0.48(1) &0.326(14) \\
$R_{\rm P}($R$_{\odot}$)&1.66(1) &1.372(27) \\
$R_{\rm S}($R$_{\odot}$)&0.96(1) &0.775(7)\\
$L_{\rm P}($L$_{\odot}$)&4.43(1) &3.531(489) \\
$L_{\rm S}($L$_{\odot}$)&0.97(1) &0.760(82) \\
\hline
\end{tabular}
\end{center}
\end{table}

\section{Various relations of contact binaries}
\label{sect_rela}
\subsection{Period-mass relation for contact binaries}
Various studies suggest that there exists a period-mass relation for contact binaries 
\citep{qian03, gaze08, gaze06_mass}. \citet{qian03} noted that the mass of the 
primary component (more massive) increases linearly with increasing period. 
\citet{gaze06_mass} also noted the same and found that the mass of the 
secondary component (less massive) is nearly period independent and varies 
between 0 and 1 $M_{\odot}$. From a sample of 112 binaries (52 W-type and 
60 A-type), \citet{gaze06_mass} derived the following relations between the 
masses of the components and the orbital period $(P)$:
\begin{equation}
\log M_{\rm P} = (0.755 \pm 0.059) \log P + (0.416 \pm 0.024),
\label{gaze_pri}
\end{equation}   
\begin{equation}
\log M_{\rm S} = (0.352 \pm 0.166) \log P - (0.262 \pm 0.067).
\label{gaze_sec}
\end{equation}    
  
\begin{figure*}
\vspace{0.03\linewidth}
\begin{tabular}{cc}
\vspace{+0.03\linewidth}
\resizebox{0.48\linewidth}{!}{\includegraphics*{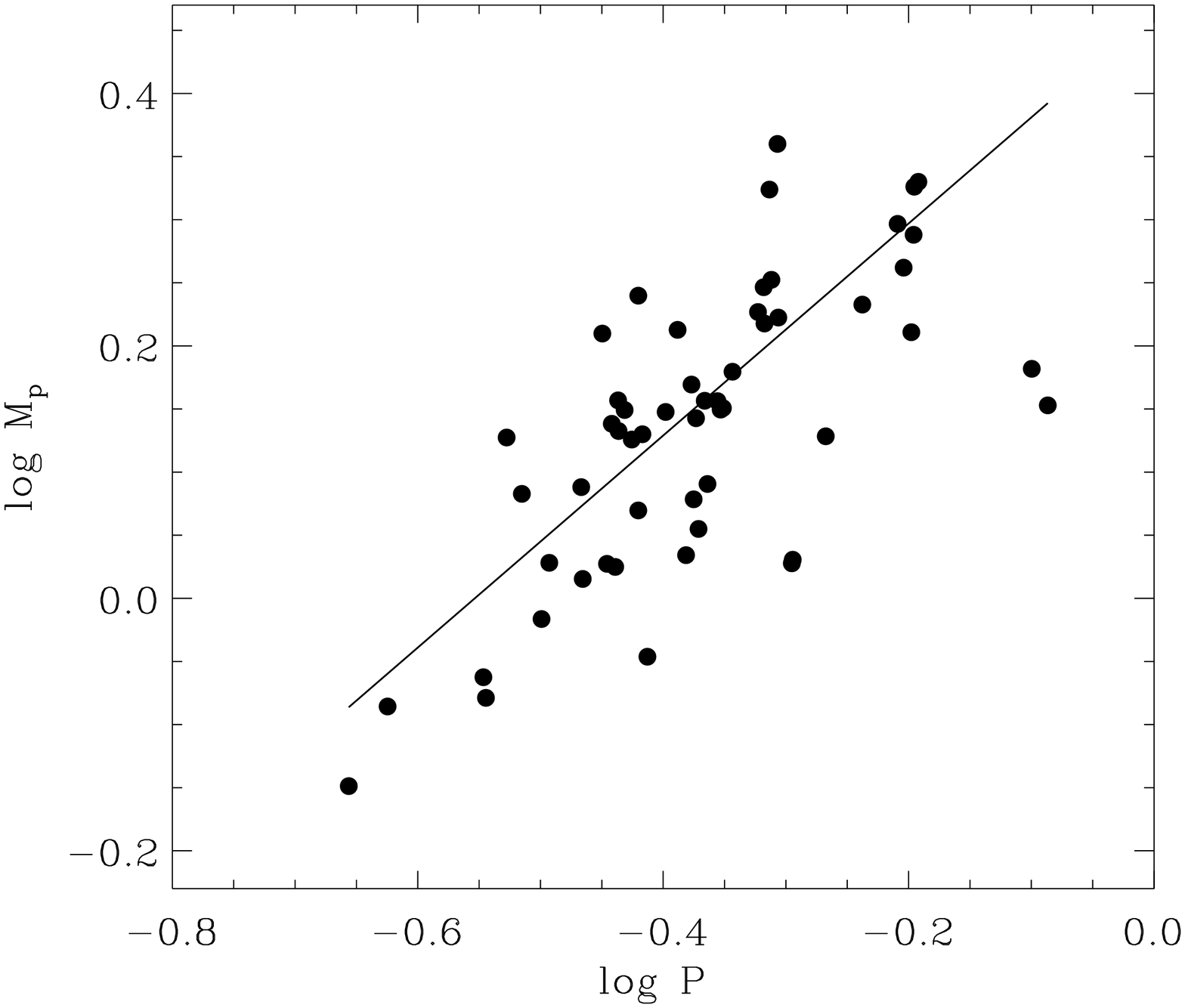}}
\resizebox{0.48\linewidth}{!}{\includegraphics*{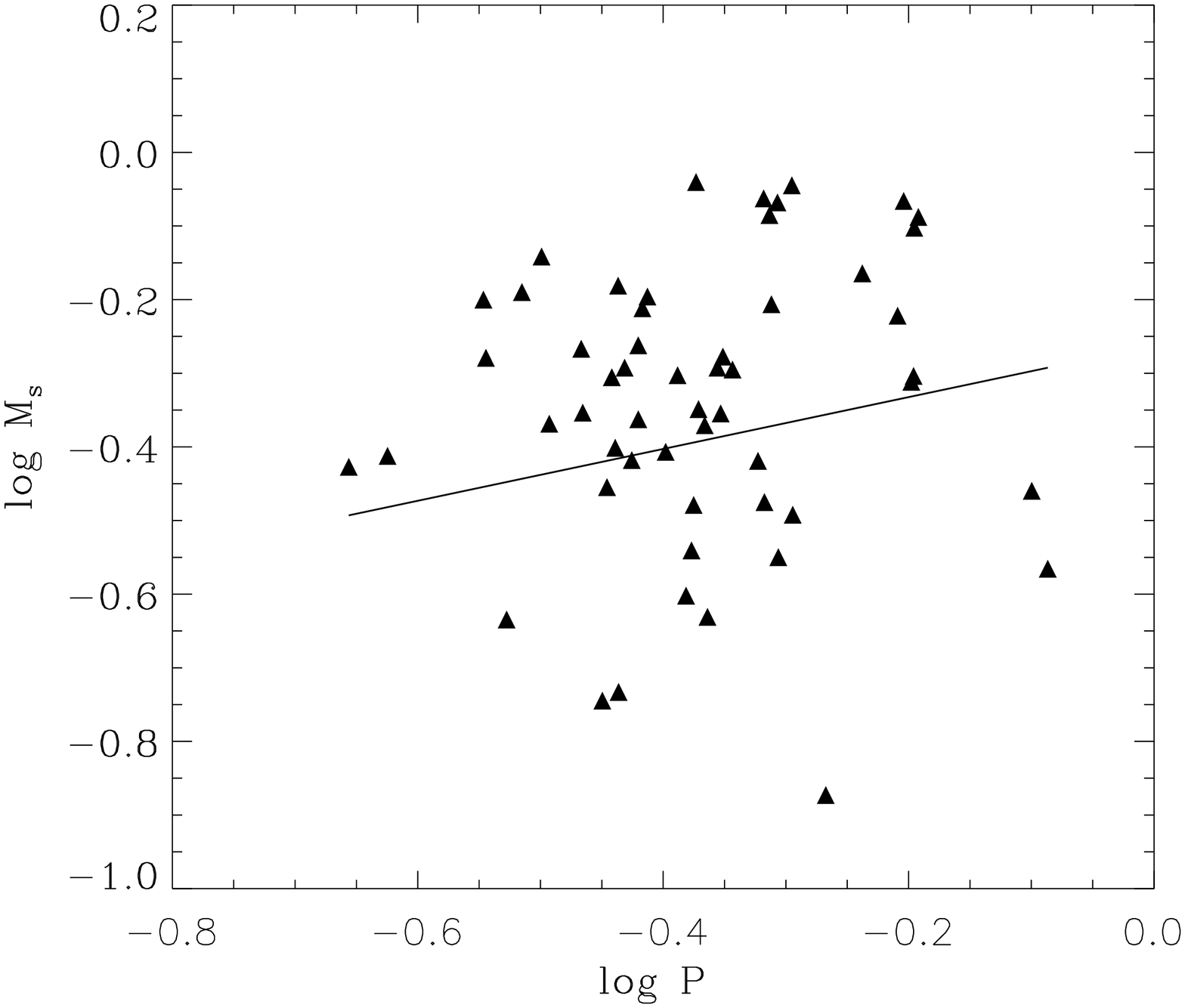}}
\vspace{-0.04\linewidth}
\end{tabular}
\caption{ The left and right panels show the masses of the primary and 
secondary components of 54 contact binaries in our sample on the 
$\log P$-$\log M_{\rm P}$ and $\log P$-$\log M_{S}$ plane. Solid lines in 
the two plots are the linear fits from equations (\ref{gaze_pri}) and 
(\ref{gaze_sec}) as derived by \citet{gaze06_mass}.}   
\label{pri_sec}
\end{figure*}

In Fig.~\ref{pri_sec}, we have plotted $\log M_{\rm P}$ versus  $\log P$ and 
 $\log M_{\rm S}$ versus $\log P$ for the 54 contact binaries in our sample. 
Also overplotted are the solid lines from the relations of equations 
(\ref{gaze_pri}) \& (\ref{gaze_sec}) in this paper obtained by 
\citet{gaze06_mass}. It can be seen that the masses of the primary components, 
in general, obey the period-mass relation defined by 
equation~(\ref{gaze_pri}), whereas, the secondary components do not seem to 
obey any strict period-mass relation of equation~(\ref{gaze_sec}) as derived 
by \citet{gaze06_mass}.   
    
\subsection{Period-color relations of contact binaries}
\begin{figure*}
\begin{center}
\includegraphics[width=8cm,height=8cm]{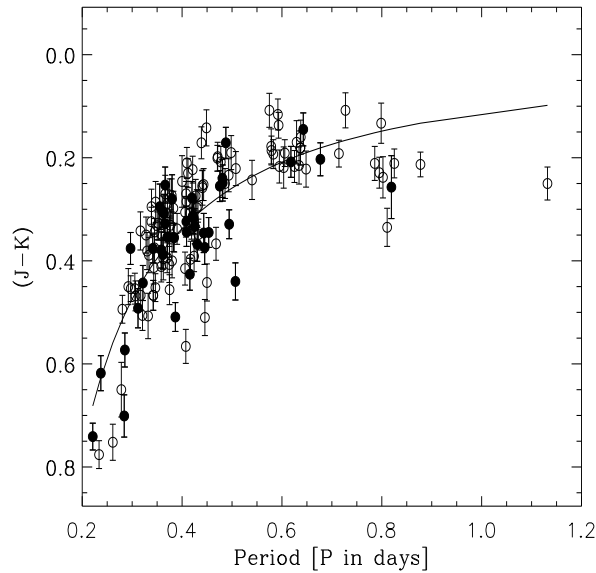}
\end{center}
\caption{Period($P$)-color $(J-K)$ relation of 141 contact binaries 
obtained from the combined data of 54 ASAS contact binaries and data from 
\citet{bili05}. The 39 stars common to our sample are shown by filled circles. 
Solid line is the power law fit to the data of these contact binaries as given 
in equation~(\ref{jhk}).}   
\label{period_color_fig}
\end{figure*}
Contact binaries obey a well-defined period-color relation. This relation 
plays a special role in the study of their properties \citep{ruci97a, ruci00}. 
The period-color relations of equal mass marginal contact binaries over a wide 
range of age and metallicity were studied by \citet{rube01}. \citet{ruci00} 
derived the shapes of short period blue envelopes (SPBEs) for disk systems by 
the following two power law relations:
\begin{eqnarray}
(B-V) = 0.04 P^{-2.25} 
\end{eqnarray}  
and
\begin{eqnarray}
(V-I) = 0.053 P^{-2.1} 
\end{eqnarray}
Here we examine the period $(P)$-infrared color $(J-K)$ relation of 141 
contact binaries. The relation that we derive here is just a power law fit to 
the data in the $P-(J-K)$ plane, whereas Rucinski's SPBE is rather a limit for 
color and period of stars. Because of the advantage of $(J-K)$ color not effected by 
the interstellar extinction and uncertain reddening corrections as compared to
the $(B-V)$ color, the period-color relation using the infrared color $(J-K)$ 
will be helpful for larger databases, where the period can be determined 
easily with greater accuracy. In the present study, the infrared color is 
obtained from the $JHK$ magnitudes in the 2MASS point source catalog 
\citep{cutr03}. To increase the number of samples to derive the $P$-$(J-K)$ 
relation, we took 129 W UMa contact binaries from \citet{bili05}. For one 
star, GZ And, in the sample, there is no reliable $JHK$ magnitude available in 
the 2MASS catalog. On the other hand, two stars in the Bilir's sample, V753 
Mon and HT Vir turned out to be semi-detached binaries in our analysis. 
Therefore, these three stars were rejected for deriving the period-infrared 
color relation of contact binaries. Hence, out of 129 stars in the Bilir's 
sample, we have selected 126 stars for the analysis. On the other hand, among 
126 stars, 39 of them are common to our sample of 54 contact binaries. 
Therefore, the total sample used for deriving the $(J-K)$ vs $P$ relation for 
contact binaries is 141. Fig.~\ref{period_color_fig} shows the period-color 
$(J-K)$ for all the 141 contact binaries in the total sample. From the figure, 
it is very clear that contact binaries definitely obey a tight period 
($P$)-color $(J-K)$ relation. From the 141 contact binaries, we derived the following 
relation 
\begin{eqnarray}
(J-K) = (0.11 \pm 0.01) P^{-1.19 \pm 0.08} 
\label{jhk}
\end{eqnarray}
The fit to this derived relation is shown by solid line in 
Fig.~\ref{period_color_fig}. Derivation of such an empirical period-color 
relation of contact binaries using accurate infrared colors will be very 
useful to display the properties of contact binaries discovered in large 
photometric surveys. However, the derivation of precise period-color relations 
of contact binaries need highly accurate age and metallicity estimations 
\citep{rube01}.
\subsection{M-R and M-L diagrams}
\begin{figure*}
\vspace{0.02\linewidth}
\begin{tabular}{cc}
\vspace{0.01\linewidth}
 \resizebox{0.50\linewidth}{!}{\includegraphics*{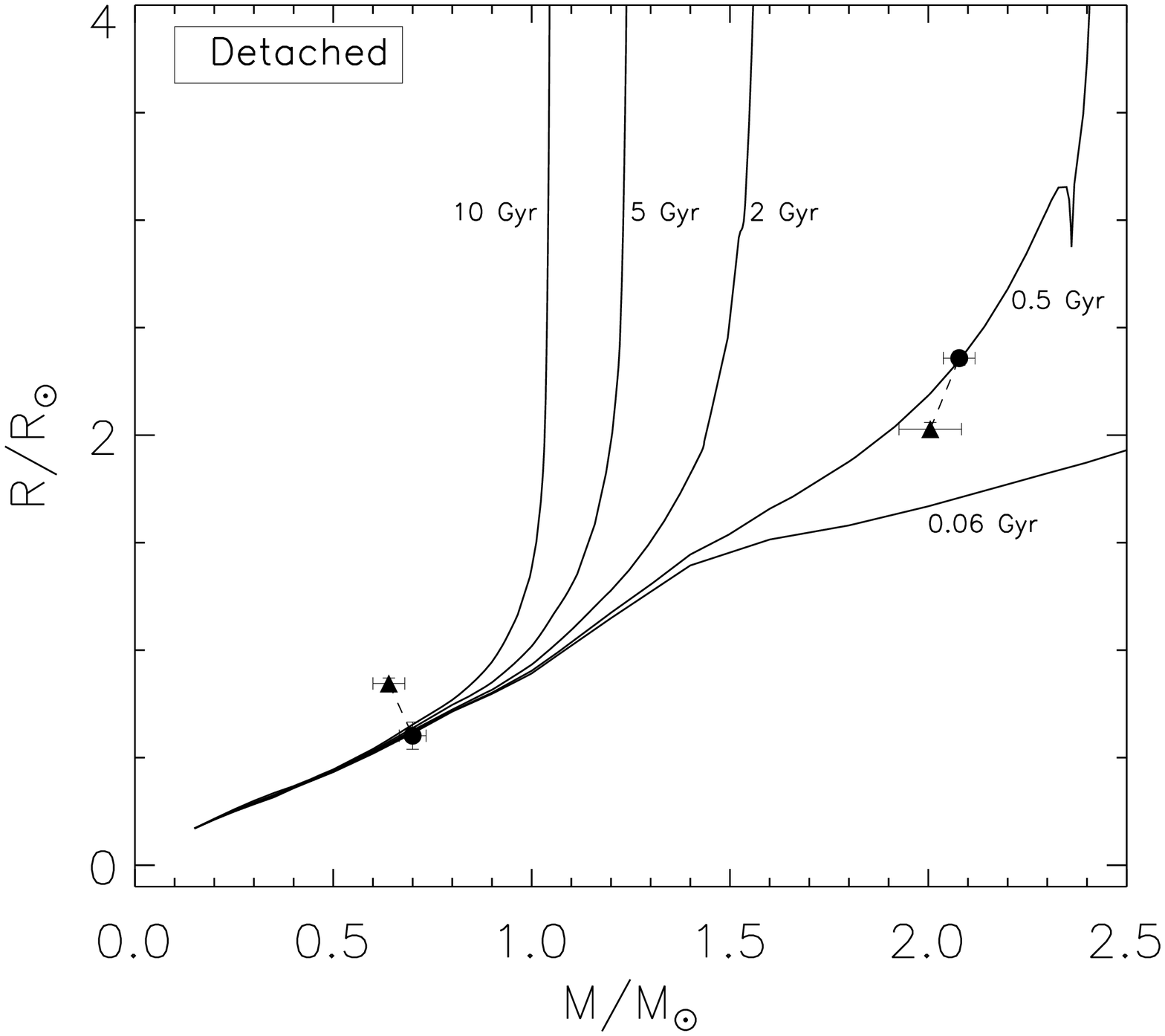}}&
  \resizebox{0.50\linewidth}{!}{\includegraphics*{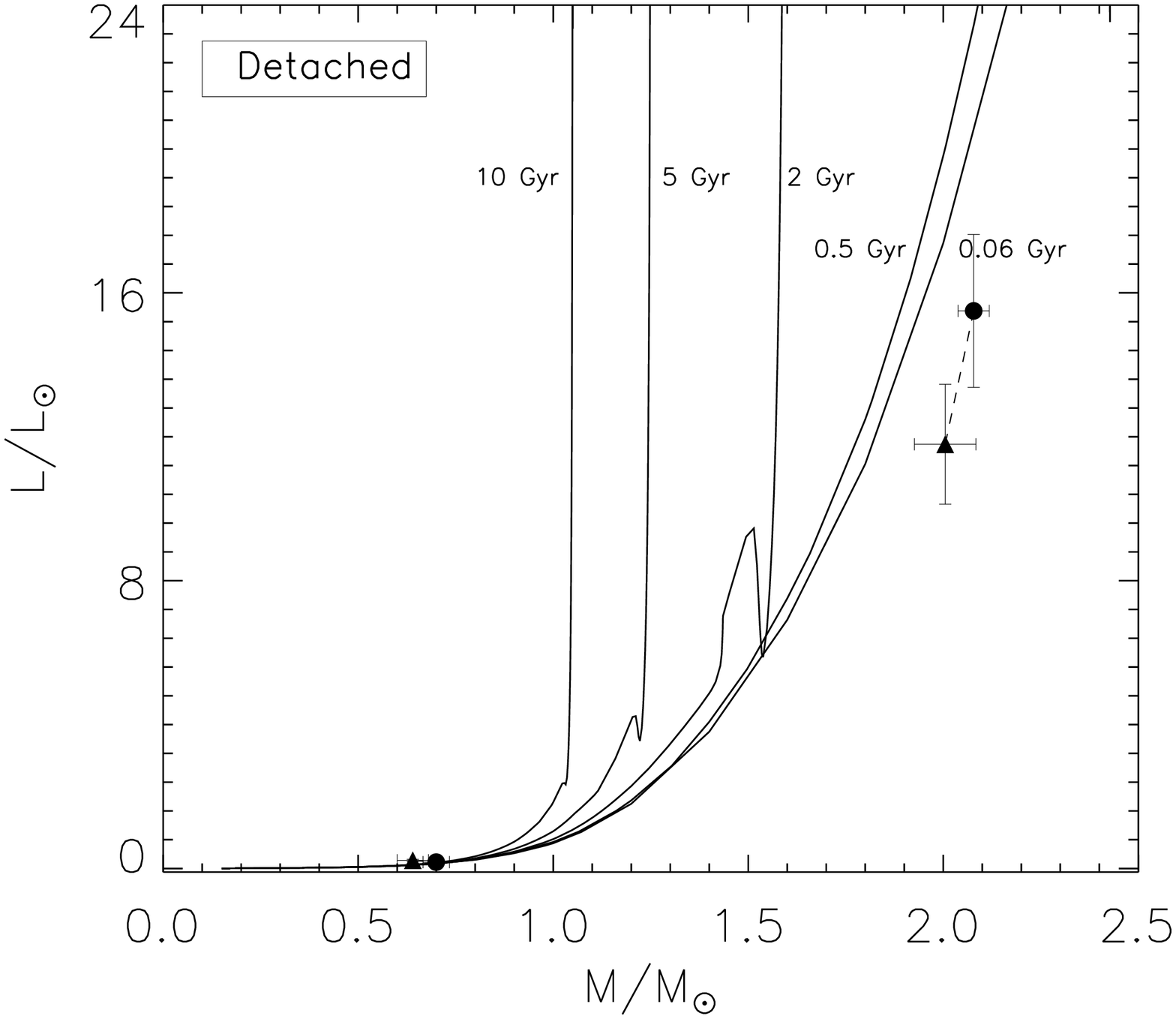}}\\
\vspace{-0.04\linewidth}
\end{tabular}
\caption{Location of the components of detached binaries on the theoretical mass-luminosity (M-L) and mass-radius 
(M-R) diagrams along with the 0.06, 0.50, 2.0, 5.0 and 10 Gyr theoretical 
isochrones from \citet{gira00} for the population I stars with the 
solar composition (X, Y, Z) = 0.708, 0.273, $0.019$. Each isochrone is plotted 
as a solid line and labelled with its corresponding age. Primary (more massive)
 components are plotted with filled circles and secondary components (less 
massive) with filled upper triangles. The components of the same pair are 
connected with dotted lines.}
\label{locations}
\end{figure*}
Fig.~\ref{locations} shows the location of the components of detached 
binaries on the theoretical mass-radius (M-R),mass-luminosity (M-L) diagrams 
along with the 0.06, 0.50, 2.0, 5.0 and 10 Gyr theoretical isochrones from 
\citet{gira00} for the population I stars with the solar composition (X, Y, Z) 
= 0.708, 0.273, $0.019$. Each isochrone is plotted as a solid line and labelled
 with its corresponding age. Primary (more massive) components are plotted 
with filled circles and secondary components (less massive) with filled upper 
triangles. The components of the same pair are connected with dotted lines. We 
have not tried to estimate the ages of the components of contact binaries. The 
ages of the primary components can be determined by interpolation from the 
isochrone fitting. As in general, to an approximation, they obey a normal 
mass-radius relations for main sequence stars. On the other hand, the 
secondary components of contact binaries are always overluminous and oversized 
for their mass and do not obey the normal main sequence mass-radius relations. 
Hence it is difficult to calculate their age from the main sequence isochrone 
fitting. However, the peculiarly evolutionary status of the contact binaries 
due to the effect of vast amount of  mass transfer is also seen. The result of 
such effect is the underluminous and undersized primary component and 
overluminous and oversized secondary component \citet{zhan09}. In fact it 
would be wrong to make comparisons between the empirical data on mass and 
absolute dimensions for such systems and the theoretical stellar evolution of 
single stars alone. The stellar evolution theory is very much complicated 
which takes into account the mass loss, the existence of the upper limit of 
the volume of a star and to allow for changes in orbital period and sizes due 
to the effect of mass transfer for semidetached and contact binaries and is 
beyond the scope of this present work. For the two detached binaries in our 
sample, 203113$+$0513.2 appears to have an age $\ge$ 10 Gyr, whereas, 
084108$-$3212.1 has an age $\le$ 0.5 Gyr from the M-R diagram.
However, determination of their accurate ages need parameter estimations 
from highly precise light curves.
   
\section{Summary and Conclusions}
\label{summary}
We have explored the ASAS database to study the eclipsing binaries. We have 
presented a detailed analysis of the light curves of 62 eclipsing binaries 
monitored by the ASAS project which have precise radial velocity 
measurements in the literature. 

We updated the ASAS periods using the ME method and classified the stars 
based on their light curve shapes. With the improved ASAS periods, we obtained
significant improvement in the light curve shapes. For preliminary 
classification, we have used cosine decomposition of the light curves into 
the Fourier coefficients, where contact, semidetached and detached 
configurations can be distinguished in the a$_{2}-$a$_{4}$ plane. 
Further, using the Roche lobe geometry, we found that the sample contains 54 
contact, 6 semidetached and 2 detached binaries. The physical parameters of 
all the 62 binaries were calculated using the WD light curve modelling 
technique. Spectroscopic mass ratios of these binaries available in the 
literature were used for their modelling. Out of 62 stars in the sample, 
light curve analysis of 39 stars are presented here for the first time using 
combined photometric and spectroscopic data. For majority of remaining 23 
stars, the determination of parameters are found to be consistent with the 
earlier values obtained in the literature. Thus, ASAS-like databases can, in 
general, be used for parameter determinations of newly discovered eclipsing 
binaries for statistical analysis. ASAS-like database opens a new avenue to 
study the parameters of eclipsing binaries in large numbers. Availability of data of a huge number of eclipsing binaries in 
these databases will facilitate for measurements of their geometrical and 
physical parameters, if precise mass ratios are determined from spectroscopy. 
Spectroscopic measurements combined with the photometry from large databases 
will provide an opportunity to understand the origin and evolution of binaries 
accurately and will resolve the long-standing issues of formation and 
evolution of contact binaries. It is therefore an urgent need to come up with 
projects which will be solely devoted to radial velocity measurements of 
eclipsing binaries discovered in various automated photometric surveys 
and hence enhance our knowledge about the formation, structure and evolution of
stars.        

\section*{Acknowledgments}
SD thanks CSIR, Govt. of India, for a Senior Research Fellowship. 
Authors acknowledge helpful discussions with Philippe Prugniel and 
Biman Jyoti Medhi. The authors thank the anonymous referee for many 
insightful comments and suggestions which have led to the substantial 
improvement of the paper. The authors thank Alceste Z. Bonanos for a discussion
on error estimation through an e-mail. The use of the SIMBAD and ADS databases 
is gratefully acknowledged. This publication makes use of data products from 
the Two Micron All Sky Survey, which is a joint project of the University of 
Massachusetts and the Infrared Processing and Analysis Center/California 
Institute of Technology, funded by the National Aeronautics and Space 
Administration and the National Science Foundation.
\bibliographystyle{mn2e}
\bibliography{deb}
\end{document}